\documentstyle[12pt]{article}
\begin{document}
\def\beq{\begin{equation}}
\def\eeq{\end{equation}}
\def\beqa{\begin{eqnarray}}
\def\eeqa{\end{eqnarray}}
\def\ra{\rightarrow}
\def\Re{{\cal R \mskip-4mu \lower.1ex \hbox{\it e}\,}}
\def\Im{{\cal I \mskip-5mu \lower.1ex \hbox{\it m}\,}}
\def\be{\begin{equation}}
\def\ee{\end{equation}}
\def\ra{\rightarrow}

\begin{titlepage}
\vspace*{-1cm}
\noindent
\phantom{bla}
\hfill{$\scriptstyle{\rm FERMILAB-Pub-94/412-T} \atop{\scriptstyle
{{\rm UMHEP-415}}
\atop{\scriptstyle {\rm SLAC-PUB-6692} \atop{\scriptstyle
{\rm UH-511-811-94}}}}$}
\\
\vskip 2.5cm
\begin{center}
{\Large\bf Radiative Weak Decays of Charm Mesons} \\
\end{center}
\vskip 1.6cm
\begin{center}
{\large Gustavo Burdman$^{(a)}$, Eugene Golowich$^{(b)}$} \\
{\large JoAnne L. Hewett$^{(c)}$ and Sandip Pakvasa$^{(d)}$} \\
\vskip .3cm
$^{(a)}$Fermilab \\
Batavia, IL 60510, USA\\
$^{(b)}$Department of Physics and Astronomy \\
University of Massachusetts, Amherst MA 01003, USA\\
$^{(c)}$Stanford Linear Accelerator Center \\
Stanford CA 94309, USA\\
$^{(d)}$Department of Physics and Astronomy \\
University of Hawaii, Honolulu HI 96822, USA \\
\vskip .3cm
\end{center}
\vskip 2cm
\begin{abstract}
\noindent
We address Standard Model predictions for flavor-changing radiative
transitions of the pseudoscalar charm mesons. Short-distance
contributions in $D$ radiative transitions are contrasted with
those in $B$ decays.  A full analysis is presented of the
$c\to u+\gamma$ electromagnetic penguin amplitude with QCD
radiative corrections included.  Given the importance of long-range
effects for the charm sector, special attention is paid to
such contributions as the vector dominance and pole amplitudes.
A number of two-body final states in exclusive charm radiative decays is
considered and the corresponding branching ratio predictions are given.
\end{abstract}
\vfill
\end{titlepage}
\vskip2truecm

\section{\bf Introduction}
Important milestones in the study of the $b$-quark system were reached
with the recent observations of both the exclusive decay
$B \to K^* \gamma$$^{\cite{cleo1}}$
\beq
{\rm B}_{B \to K^* \gamma} = (4.5 \pm 1.5 \pm 0.9 )\times 10^{-5}\ \ ,
\label{excl}
\eeq
and of the inclusive transition$^{\cite{cleoinc}}$
\beq
{\rm B}_{b \to s \gamma} = (2.32 \pm 0.57 \pm 0.35 )\times 10^{-4} \ \ .
\label{incl}
\eeq
To first approximation, these flavor-changing radiative decays can be
interpreted at the quark level in terms of the $b \to s \gamma$ transition.
The Standard Model allows for such a process by means of a one-loop
penguin-type amplitude.  Within errors, agreement of the measured branching
ratios and Standard Model predictions appears to be reasonable.
The small magnitudes of these branching ratios indicate just how sensitive
experimental probes of $b$-quark hadrons have become.  However, an
outstanding question is the size of the long distance
contributions\cite{gp,soni} to such
radiative $B$ decays relative to the short distance penguin amplitude.  This
issue must be addressed in order to establish the viability of determining the
value of the ratio of CKM matrix elements, $|V_{td}|/|V_{ts}|$, from a
measurement of the ratio of exclusive branching fractions
$B_{B\to\rho\gamma}/B_{B\to K^*\gamma}$.

Important as they are, the above measurements by no means exhaust the
set of interesting problems.  It has become increasingly evident that
the database for charm hadrons is also in a state of rapid expansion, and that
physically important levels of sensitivity are being achieved.  Perhaps
the most impressive example of this to-date is the recent observation of
the nonleptonic decay $D^0 \to K^+ \pi^-$, with branching
ratio$^{\cite{cleo2}}$
\beq
{{\rm B}_{D^0 \to K^+ \pi^-} \over
{\rm B}_{D^0 \to K^- \pi^+}}
= 0.0077 \pm 0.0025 \pm 0.0025 \ \ .
\label{dcs}
\eeq
This transition has been interpreted as evidence of a doubly Cabibbo
suppressed transition rather than of $D^0$-${\bar D}^0$ mixing.

The discussion in this paper will be directed towards a somewhat different
aspect of charm physics, the flavor-changing radiative decays.  These
transitions require the joint occurrence of weak and electromagnetic
interactions.  From Table~1$^{\cite{pdg},\cite{selen},\cite{fte}}$, we see
that no such events (involving emission of real or virtual photons) have
yet been observed.  However, these decays are an active area of study, and
data gathered in ongoing fixed-target experiments are establishing markedly
improved bounds.  Our objective in the analysis to follow will be
to provide up-to-date predictions for flavor-changing radiative
transitions of charm systems.  Since the experimental
situation for charm mesons is at present more favorable than for
charm baryons, we shall restrict our attention to the former.  Even with
this restriction, it is a tall order to supply accurate theoretical values.
It has become evident over a long period of time that theoretical calculations
of $D$-meson weak decays are not particularly trustworthy, due in part to the
absence of a rapidly convergent approximation scheme and also to the
presence of significant hadron dynamical effects in the $D$ meson mass region.
Despite this, we feel that one can make some definite statements, such
as the relative importance of long-range and short-range effects and of
the various types of final states which can reasonably be anticipated.
We shall base our analysis on a variety of theoretical techniques,
from operator-product expansion and renormalization-group methods
to more phenomenological approaches like vector-meson-dominance (VMD).
Measurement of radiative charm decays would probe the long distance
contributions and thus provide further insight
in the extrapolation of calculational techniques to the $B$ sector.
\phantom{xxxx}\vspace{0.1in}
\begin{center}
\begin{tabular}{lc}
\multicolumn{2}{c}{Table~1 {Status of Electroweak-induced Charm Decays}}\\
\hline\hline
Mode & Branching Ratio \\ \hline
$D^0 \to \rho^0 \gamma$ & $<1.4\times 10^{-4}$ \\
$D^0 \to \phi^0 \gamma$ & $<2.0\times 10^{-4}$ \\
$D^0 \to {\bar K}^{0} e^+ e^-$ & $<1.7\times 10^{-3}$ \\
$D^0 \to \rho^0 e^+ e^-$ & $< 4.5 \times 10^{-4}$ \\
$D^0 \to \rho^0 \mu^+ \mu^-$ & $< 8.1 \times 10^{-4}$ \\
$D^+ \to \pi^+ e^+ e^-$ & $< 2.5 \times 10^{-3}$ \\
$D^+ \to \pi^+ \mu^+ \mu^-$ & $< 2.9 \times 10^{-3}$ \\
$D^+ \to K^+ e^+ e^-$ & $< 4.8 \times 10^{-3}$ \\
$D^+ \to K^+ \mu^+ \mu^-$ & $< 9.2 \times 10^{-3}$ \\ \hline\hline
\end{tabular}
\end{center}
\phantom{xxxx}\vspace{0.1in}

Let us summarize the contents to follow.  In Section~2, we consider the
short-range component in radiative charm decays, primarily
the charm counterpart of the penguin amplitude which dominates
the radiative $B$-meson decays.  In addition to addressing
$c$-quark physics, our analysis contains a purely theoretical
advance by removing an unnecessary assumption made in earlier studies
involving $b$-quark applications.  Section~3 begins our analysis of the
so-called `long-distance' contributions with an analysis of pole
diagrams, which are induced by the weak mixing of pseudoscalar
and/or vector charm mesons with noncharm states.  In Section~4, we
continue our study of long distance effects by turning our attention
to a study of VMD amplitudes.  Our conclusions and recommendations
for future studies are given in Section~5.  There is also an Appendix
in which the applicability of VMD to certain light-meson decays is
commented on.

\section{\bf Short Distance Contributions}
Examples of diagrams which mediate the
short-distance transition amplitudes for radiative charm decay are depicted
in Fig.~1.  They have in common that the photon emission occurs
in a region of spacetime determined by the propagator of the $W$-boson.
In view of the large $W$-mass $M_{\rm W}$, this region has a very limited
extent compared to the length scale of the strong interactions, hence
the name `short-distance'.  Looking
ahead, our conclusion regarding such short-distance
amplitudes will be that in radiative decays of charm mesons they are
small relative to long-distance effects, even though they receive large
enhancements from QCD corrections.  As described earlier,
this is of course in stark contrast to $B$ decay.

\phantom{xxxx}\vspace{0.03in}
\begin{center}
\begin{tabular}{c}\phantom{xxxxxxxxxxxxxxxxxxxxxx} \\
\phantom{xxxxxxxxxxxxxxxxxxxxxx} \\ \hline
\phantom{xxxxxxxxxxxxxxxxxxxxxx} \\
\phantom{xxxxxxxxxxxxxxxxxxxxxx} \\
\phantom{xxxxxxxxxxxxxxxxxxxxxx} \\
\phantom{xxxxxxxxxxxxxxxxxxxxxx} \\ \hline
\phantom{xxxxxxxxxxxxxxxxxxxxxx} \\
\phantom{xxxxxxxxxxxxxxxxxxxxxx} \\
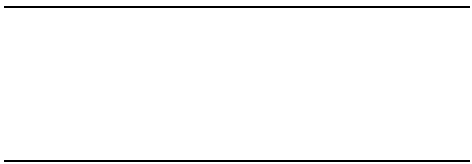
{Figure 1. Short-distance Effects} \\
\end{tabular}
\end{center}
\vspace{0.08in}

In recognition of the importance attached to the electromagnetic
penguin transition in radiative $B$ decays, we give a brief pedagogical
comparison between the role played by this effect for $B$ and for $D$
decay.  The two transitions in question are given at the quark-level by
\beqa
b ({\bf p},\lambda ) &\to& s ({\bf p}',\lambda ') +
\gamma ({\bf q},\sigma ) \nonumber \\
c ({\bf p},\lambda ) &\to& u ({\bf p}',\lambda ') +
\gamma ({\bf q},\sigma ) \label{cu} \ \ .
\eeqa
To highlight the crucial role played by the quark masses and CKM
matrix elements, let us at first ignore the effect of QCD radiative
corrections.  The relevant Feynman diagrams are then depicted in
Figs.~1(a),1(c) and the penguin amplitude for the transition of a heavy
quark $Q$ to a much lighter quark $q$ and an on-shell photon is
given by$^{\cite{il}}$
\beqa
 \lefteqn{{\cal A}^{\rm (EM~peng)}_{Q \to q \gamma} =} \label{il1} \\
& & {eG_{\rm F}\over 4\sqrt{2}\pi^2} \sum_i \lambda_i F_2(x_i)
{}~{\bar u}_q (p',\lambda ' )\epsilon^{\mu\dagger} (q,\lambda )
\sigma_{\mu\nu}q^\nu [m_QP_R+m_qP_L]u_Q (p ,\lambda) \ , \nonumber
\eeqa
where $P_{R(L)}$ are the right(left)-handed helicity projection operators,
$x_i \equiv {m_i^2 / M_{\rm W}^2}$, $\lambda_i\equiv {V_{\rm is}}^*V_{\rm ib}$
for ${\cal A}_{b\to s\gamma}$ and $\lambda_i\equiv {V_{\rm ci}}^*V_{\rm ui}$
for ${\cal A}_{c\to u\gamma}$.  The function $F_2$ gives the contribution of
each internal quark to the electromagnetic penguin loop,
\beqa
\lefteqn{F_2(x)  =} \nonumber \\
& & Q \left[ {x^3-5x^2-2x\over 4(x-1)^3}+
{3x^2\ln x\over 2(x-1)^4}\right]
+ {2x^3+5x^2-x\over 4(x-1)^3} -{3x^3\ln x\over 2(x-1)^4} \,,
\label{il2}
\eeqa
with $Q$ being the charge of the internal quark.
For $b\to s \gamma$ the sum is carried out over the quarks $u,c,t$ and
the term proportional to the $s$-quark mass in Eq.~(\ref{il1}) is generally
neglected, whereas for $c \to u \gamma$, one sums over the quarks $d , s , b$
and ignores the corresponding term proportional to the $u$-quark mass.

Let us get acquainted with some of the numerical values.
In Table~2, we first display the magnitude of the function $F_2$ and
then fold in the CKM dependence for the $b\to s \gamma$ transition (we
take $m_u=5$~MeV, $m_c=1.5$~GeV, $m_t=174$~GeV, and the central values of
the CKM matrix elements as given in Ref.~\cite{pdg}).
\phantom{xxxx}\vspace{0.1in}
\begin{center}
\begin{tabular}{ccc}
\multicolumn{3}{c}{Table~2 {Contributions to $b\to s + \gamma$}}\\ \hline\hline
Quark & $F_2$ & $|V_{\rm ib}^{\phantom{x}} V_{\rm is}^{~*}| F_2$  \\ \hline
$u$ & $2.27\times 10^{-9}$ & $1.29\times 10^{-12}$ \\
$c$ & $2.03\times 10^{-4}$ & $7.34\times 10^{-6}$ \\
$t$ & $0.39$ & $1.56\times 10^{-2}$ \\ \hline\hline
\end{tabular}
\end{center}
\phantom{xxxx}\vspace{0.1in}
Dominance of the $t$-quark intermediate state is evident, even upon including
the CKM factors.  Its effect is so large that the other intermediate states
are numerically negligible and hence are typically omitted.
The corresponding situation is given for $c \to u \gamma$ in Table~3 (with
$m_d=11$~MeV, $m_s=150$~MeV, and $m_b=4.9$~GeV).
\phantom{xxxx}\vspace{0.08in}
\begin{center}
\begin{tabular}{ccc}
\multicolumn{3}{c}{Table~3 {Contributions to $c \to u + \gamma$}}\\
\hline\hline
Quark & $F_2$ & $|V_{\rm ci}^{~*} V_{\rm ui}| F_2$  \\ \hline
$d$ & $1.57\times 10^{-9}$ & $3.36\times 10^{-10}$ \\
$s$ & $2.92\times 10^{-7}$ & $6.26\times 10^{-8}$ \\
$b$ & $3.31\times 10^{-4}$ & $3.17\times 10^{-8}$ \\ \hline\hline
\end{tabular}
\end{center}
\phantom{xxxx}\vspace{0.08in}
The amplitude for $c\to u\gamma$ differs from that of $b\to s\gamma$ in
two important respects, (i) there is no single intermediate state which
dominates, and (ii) the overall magnitude is much smaller.

Neglecting the final state fermion mass,
the QCD uncorrected decay rate $\Gamma^{(0)}_{Q\to q\gamma}$ is given by
\beq
\Gamma^{(0)}_{Q\to q\gamma} = {\alpha G_F^2\over 128 \pi^4}m_Q^5
\bigg| \sum_i \lambda_i F_2(x_i)\bigg|^2 \,.
\label{il4}
\eeq
To obtain the branching fraction, the inclusive rate is scaled to that of
the semi-leptonic decay $Q\to q'\ell\nu$.  This procedure removes
uncertainties in the calculation due to the overall factor of $m_Q^5$
which appears in both expressions, and reduces the ambiguities involved
with the imprecisely determined CKM factors.  Taking the above numerical
values for the internal quark masses, and using the values of the
semi-leptonic branching ratios as given in Ref.~\cite{pdg}, this yields
\beqa
{\rm B}_{b\to s\gamma} & = & {3\alpha\over 2\pi}\cdot {|V_{tb}V^*_{ts}
F_2(x_t)|^2\over |V_{cb}|^2\left[ g(m_c/m_b)+{ \mbox{$|V_{ub}|^2$}\over
\mbox{$|V_{cb}|^2$}} g(m_u/m_b) \right] } \cdot {\rm B}_{B\to X\ell\nu}
\,, \nonumber \\
& = & 1.29\times 10^{-4} \,, \\
{\rm B}_{c\to u\gamma} & = & {3\alpha\over 2\pi}\cdot {|V^*_{cs}V_{us}F_2(x_s)
+V^*_{cb}V_{ub}F_2(x_b)|^2\over |V_{cs}|^2\left[ g(m_s/m_c)+
{\mbox{$|V_{cd}|^2$}\over \mbox{$|V_{cs}|^2$}} g(m_d/m_c)\right] }
\cdot {\rm B}_{D^+\to X\ell^+\nu}
\,, \nonumber \\
& = & 1.39\times 10^{-17} \,. \nonumber
\label{br0}
\eeqa
Here, the function $g(x)$ is the usual phase space factor in semi-leptonic
meson decay, where constituent values of the final-state quark masses have
been used.$^{\cite{cab}}$  The QCD uncorrected $c \to u\gamma$ transition
is seen to have an unobservably small branching fraction.

We next examine the impact of the QCD radiative corrections on
the above branching ratios.  We begin by reviewing the calculation for
the $b\to s \gamma$ transition, which will serve as the foundation of our
subsequent discussion of $c\to u\gamma$.  The QCD corrections are
calculated$^{\cite{{qcd},{gsw}}}$ via an operator product expansion
based on the effective hamiltonian
\beq
H^{|\Delta b|=1}_{\rm eff}=-{4G_F\over\sqrt 2}\lambda_t
\sum_{k=1}^8c_k(\mu)O_k(\mu) \,,
\label{effh}
\eeq
where the $\{ O_k\}$ are a complete set of renormalized dimension-six
operators involving light fields which govern the $b\to s$ transitions.
They consist of two current-current operators $O_{1,2}$, four strong penguin
operators $O_{3-6}$, and electro- and chromo-magnetic dipole operators
$O_{7}$ and $O_8$,
\beqa
O_{1} & = & (\bar c_\alpha\gamma_\mu P_L b_\beta)(\bar s_\beta\gamma^\mu P_L
c_\alpha) \,, \nonumber \\
O_{2} & = & (\bar c_\alpha\gamma_\mu P_L b_\alpha)(\bar s_\beta\gamma^\mu
P_L c_\beta) \,, \nonumber \\
O_3 & = & (\bar s_\alpha\gamma_\mu P_L b_\alpha)\sum_q(\bar q_\beta\gamma^\mu
P_L q_\beta) \,, \nonumber \\
O_4 & = & (\bar s_\alpha\gamma_\mu P_L b_\beta)\sum_q(\bar q_\beta\gamma^\mu
P_L q_\alpha) \,, \nonumber \\
O_5 & = & (\bar s_\alpha\gamma_\mu P_L b_\alpha)\sum_q(\bar q_\beta\gamma^\mu
P_R q_\beta) \,, \label{effh1} \\
O_6 & = & (\bar s_\alpha\gamma_\mu P_L b_\beta)\sum_q(\bar q_\beta\gamma^\mu
P_R q_\alpha) \,, \nonumber \\
O_7 & = & {e\over 16\pi^2}m_b(\bar s_\alpha\sigma_{\mu\nu}P_R b_\alpha)
F^{\mu\nu} \,, \nonumber \\
O_8 & = & {g_s\over 16\pi^2}m_b(\bar s_\alpha\sigma_{\mu\nu}T^a_{\alpha\beta}
P_R b_\beta)G^{a\mu\nu} \ .  \nonumber
\eeqa
The above effective hamiltonian is then evolved from the electroweak scale
down to $\mu\sim m_b$ by the Renormalization Group Equations (RGE).

In the RG analysis, the Wilson coefficients are to be
evaluated perturbatively at the $W$ scale where the matching conditions
are imposed and then evolved down to the renormalization scale
$\mu$.  The expressions for the $\{ c_k\}$ at the $W$ scale are
\beq
\begin{tabular}{lcl}
$c_{1,3-6}(M_W)$ &=& $0$\ , \\
$c_7(M_W)$ &=& $-{1\over 2}F_2(x_t)$\ , \\
\end{tabular}
\qquad
\begin{tabular}{lcl}
$c_2(M_W)$ &=& $1$ \ ,\\
$c_8(M_W)$ &=& $-{1\over 2}D(x_t)$ \ \ .\\
\end{tabular}
\label{cimw}
\eeq
with
\beq
D(x)={x^3-5x^2-2x\over 4(x-1)^3}+{3x^2\ln x\over 2(x-1)^4} \ \,.
\label{c8mw}
\eeq
The solution to the RGE at the leading logarithmic order is given by
\beq
c_k^{eff}(\mu)=U^5_{k\ell}(\mu, M_W)c_\ell(M_W) \,,
\label{ceff}
\eeq
where $U^5_{k\ell}$ denotes the evolution matrix in a five-flavor context
and is determined by
\beq
U^5(m_1,m_2)_{kn}=O_{k\ell}\left[ \eta^{\vec{a}_\ell}\right]O^{-1}_{\ell n} \,.
\label{u5}
\eeq
In the above we define $\eta\equiv\alpha_s(m_2)/\alpha_s(m_1)$ and
$\vec{a}_\ell\equiv\gamma^D_{\ell\ell}/2\beta_0$ ({\it not} summed on $\ell$),
where $\beta_0=11-2n_f/3$ and $\gamma^D=O^{-1}\, \gamma^{(eff)\, T}\, O$ is the
diagonalized form of the $8\times 8$ anomalous dimension matrix.  We use the
scheme-independent form of the matrix $\gamma^{eff}$, which is given
explicitly in Ref.~\cite{bmmp} in terms of the number of $Q=+2/3$ and
$Q=-1/3$ quarks present in the effective theory.

Scaling again to the semi-leptonic decay, the branching fraction is now given
by
\beq
{\rm B}_{b\to s\gamma} =  {6\alpha\over \pi}\cdot \left| {V_{tb}V^*_{ts}\over
V_{cb}}\right|^2 \cdot {|c_7^{eff}(\mu)|^2\over
g(m_c/m_b)+{ \mbox{$|V_{ub}|^2$}\over \mbox{$|V_{cb}|^2$}}
g(m_u/m_b) } \cdot {\rm B}_{B\to X\ell\nu} \,,
\label{br}
\eeq
The numerical values of the separate contributions to $c_7^{eff}(\mu)$ are,
with $m_t = 174$ GeV (for illustration purposes), $\mu=m_b=4.87$ GeV, and
$\alpha_s(M_Z)=0.124$ as determined by LEP$^{\cite{lepalpha}}$,
\beqa
c_7^{eff}(\mu) & = & 0.670\, c_7(M_W) + 0.091\, c_8(M_W) - 0.172\, c_2(M_W)
\nonumber \,, \\
               & = & 0.670(-0.195) + 0.091(-0.097) - 0.172 \ \,.
\label{c7eff}
\eeqa
Taking the overall CKM factor in the branching fraction to be unity, and
$|V_{ub}|/|V_{cb}|=0.08$, this procedure yields
\begin{equation}
{\rm B}_{b\to s\gamma} = \left( 2.92^{+0.77}_{-0.59}\right) \times 10^{-4} \,.
\label{qcdbr}
\end{equation}
The central value corresponds to $\mu=m_b$, while the upper and
lower errors represent the deviation due to assuming $\mu=m_b/2$ and
$\mu=2m_b$, respectively.  We see that this value compares favorably to
the recent CLEO measurement$^{\cite{cleoinc}}$ of the inclusive rate
cited earlier in Eq.~(\ref{incl}).  When compared with the uncorrected
result of Eq.~(\ref{br0}), the QCD corrections are seen to increase the
branching ratio by roughly a factor of $2$.

We take this opportunity to reflect further on the size of the QCD corrections.
Earlier estimates$^{\cite{desh}}$ of these corrections found that the
enhancements to the $b\to s\gamma$ branching fraction were more than an
order of magnitude for $m_t<M_W$.  This is because the effect of the QCD
radiative correction to the weak vertex is to replace the GIM power
suppression in Eq.~(\ref{il1}) by a logarithmic suppression.
We explicitly illustrate this effect in Fig.~2, where we show the dependence
of the $c_7$ Wilson coefficient on the mass of a single internal quark using
the calculational procedure described above.  In the lower of the two curves,
the dependence of $c_7$ determined at scale $\mu=m_W$ is displayed, while
the upper curve corresponds to the evolved $c_7$ evaluated at $\mu=m_b$.
We see that $c_7(\mu=m_b)$ is a reasonably flat function of the
intermediate quark mass, and that the corrections are
substantial for light internal quarks, with an increase
of 3-4 orders of magnitude in the rate for $m_q=5 - 10$~GeV.
For the case of one heavy internal quark, {\it e.g.}, $b\to s\gamma$ with
$m_t>M_W$, we see that the GIM mechanism no longer plays such a crucial role
and the QCD enhancements are not as dramatic.

\phantom{xxxx}\vspace{0.03in}
\begin{center}
\begin{tabular}{c}\phantom{xxxxxxxxxxxxxxxxxxxxxx} \\
\phantom{xxxxxxxxxxxxxxxxxxxxxx} \\ \hline
\phantom{xxxxxxxxxxxxxxxxxxxxxx} \\
\phantom{xxxxxxxxxxxxxxxxxxxxxx} \\
\phantom{xxxxxxxxxxxxxxxxxxxxxx} \\
\phantom{xxxxxxxxxxxxxxxxxxxxxx} \\
\phantom{xxxxxxxxxxxxxxxxxxxxxx} \\
\phantom{xxxxxxxxxxxxxxxxxxxxxx} \\
\phantom{xxxxxxxxxxxxxxxxxxxxxx} \\
\phantom{xxxxxxxxxxxxxxxxxxxxxx} \\
\phantom{xxxxxxxxxxxxxxxxxxxxxx} \\ \hline
\phantom{xxxxxxxxxxxxxxxxxxxxxx} \\
\phantom{xxxxxxxxxxxxxxxxxxxxxx} \\
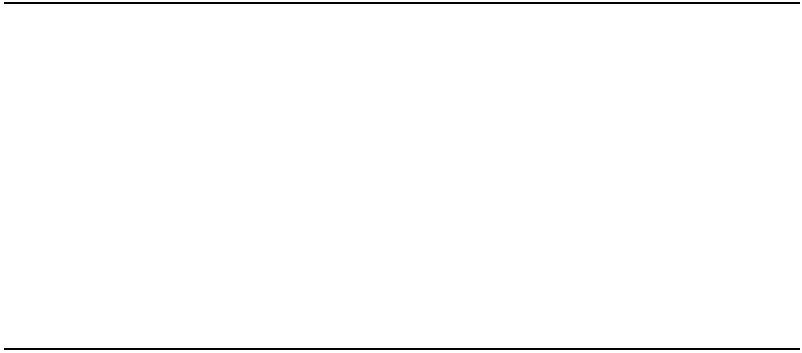
{Figure 2. Dependence of $c_7$ on Intermediate-quark Mass} \\
\end{tabular}
\end{center}
\vspace{0.08in}

We now consider the case of radiative charm transitions.  The $|\Delta c|=1$
effective hamiltonian can be written as
\beq
H^{|\Delta c|=1}_{\rm eff}={-4G_F\over\sqrt 2}\lambda_b
\sum_{k=1}^{10}c_k(\mu)O_k(\mu) \,,
\label{cheff}
\eeq
with $\lambda_i=V^*_{ci}V_{ui}$ as defined previously.  The CKM structure of
the operators differs dramatically from the $b\to s\gamma$ case.  Here,
both $O_1$ and $O_2$ have {\it two} contributions which have approximately
equal CKM weighting since $\lambda_s\simeq\lambda_d$.  We stress that
extreme caution must be exercised in order to correctly incorporate these
terms.  To be precise we explicitly separate $O_{1}$ and $O_{2}$ into
two operators according to their CKM structure,
\beqa
O_{1a} & = & (\bar u_\alpha\gamma_\mu P_Ls_\beta)(\bar s_\beta\gamma^\mu P_L
c_\alpha)\,, \quad\quad\quad
O_{1b}  =  (\bar u_\alpha\gamma_\mu P_Ld_\beta)
(\bar d_\beta\gamma^\mu P_Lc_\alpha) \,, \nonumber \\
O_{2a} & = & (\bar u_\alpha\gamma_\mu P_Ls_\alpha)(\bar s_\beta\gamma^\mu
P_Lc_\beta)\,, \quad\quad\quad
O_{2b}  =  (\bar u_\alpha\gamma_\mu P_Ld_\alpha)
(\bar d_\beta\gamma^\mu P_Lc_\beta) \,, \nonumber \\
\label{newop}
\eeqa
and write the remaining $|\Delta c| = 1$ operators in a form analogous
to their $|\Delta b| = 1$ counterparts,
\beqa
O_3 & = & (\bar u_\alpha\gamma_\mu P_Lc_\alpha)\sum_q(\bar q_\beta\gamma^\mu
P_Lq_\beta) \,, \nonumber \\
O_4 & = & (\bar u_\alpha\gamma_\mu P_Lc_\beta)\sum_q(\bar q_\beta\gamma^\mu
P_Lq_\alpha) \,, \nonumber \\
O_5 & = & (\bar u_\alpha\gamma_\mu P_Lc_\alpha)\sum_q(\bar q_\beta\gamma^\mu
P_Rq_\beta) \,, \label{effop} \\
O_6 & = & (\bar u_\alpha\gamma_\mu P_Lc_\beta)\sum_q(\bar q_\beta\gamma^\mu
P_Rq_\alpha) \,, \nonumber \\
O_7 & = & {e\over 16\pi^2}m_c(\bar u_\alpha\sigma_{\mu\nu}P_Rc_\alpha)
F^{\mu\nu} \,, \nonumber \\
O_8 & = & {g_s\over 16\pi^2}m_c(\bar u_\alpha\sigma_{\mu\nu}T^a_{\alpha\beta}
P_Rc_\beta)G^{a\mu\nu} \,, \nonumber
\eeqa
where the terms proportional to $m_u$ in $O_{7,8}$ have again been neglected.
Since the quantity $-\lambda_b$ has been factorized in Eq.~(\ref{cheff}) above,
the values of the corresponding Wilson coefficients at the matching scale are
now
\beq
\begin{tabular}{lcl}
$c_{1a}(M_W)$ &=& $0$ , \\
$c_{2a}(M_W)$ &=& $ -\lambda_s / \lambda_b$ , \\
\end{tabular}
\qquad
\begin{tabular}{lcl}
$c_{1b}(M_W)$ &=& $0$ \ ,\\
$c_{2b}(M_W)$ &=& $ -\lambda_d / \lambda_b$ \ \ .\\
\end{tabular}
\label{coeffs}
\eeq
The values of the Wilson coefficients for $c_{3-6}(M_W)$ are the same as in
Eq.~(\ref{cimw}), and the coefficients $c_{7,8}(M_W)$ are modified to
\beqa
c_7(M_W) & = & -{1\over 2}\left[{\lambda_s\over\lambda_b} F_2(x_s)+F_2(x_b)
\right] \,, \nonumber \\
c_8(M_W) & = & -{1\over 2}\left[{\lambda_s\over\lambda_b} D(x_s)+D(x_b)
\right] \ ,
\label{c7mwc}
\eeqa
with each containing intermediate $s$-quark and $b$-quark contributions.
Due to the CKM dependence, $c_{7,8}(M_W)$ now contain both real and imaginary
terms which in principle must be evolved separately.  We note that the real
parts of the $x_s$-dependent terms are numerically the same order of magnitude
as the $x_b$ terms.  Now we evolve the effective theory down to the scale
$\mu\sim m_c$.  This takes place in two successive steps; first, we go from
the electroweak scale down to $m_b$ working in an effective 5 flavor theory,
and then to $\mu<m_b$ in an effective 4 flavor theory.  This procedure is
similar to what is performed for the $|\Delta s| = 1$ kaon
transitions$^{\cite{kaon}}$, where the effective theory is evolved to
$\mu\sim 1$ GeV in 3 successive steps.  We then have
\beqa
\Re c_k^{eff}(\mu) & = & U^4_{k\ell}(\mu, m_b)U^5_{\ell n}(m_b, M_W)
\Re c_n(M_W) \,, \nonumber \\
\Im c_k^{eff}(\mu) & = & U^4_{k\ell}(\mu, m_b)U^5_{\ell n}(m_b, M_W)
\Im c_n(M_W) \,, \label{creim}
\eeqa
and
\beq
|c_7^{eff}(\mu)|^2=|\Re c_7^{eff}(\mu)|^2 + |\Im c_7^{eff}(\mu)|^2 \,.
\label{c7sq}
\eeq
The renormalization group evolution matrices $U^4$ and $U^5$ are determined
as in Eq.~(\ref{u5}) now using a $10\times 10$ anomalous dimension matrix
$\gamma^{eff}$.  We take the anomalous dimensions of the split operators
$O_{1a-b}$ and $O_{2a-b}$ to be exactly those for $O_1$ and $O_2$,
respectively,
as the anomalous dimensions do not depend on the CKM structure of the operator.
We use the form of $\gamma^{eff}$ as given in Ref.~\cite{bmmp}, taking care to
keep $n_f=4$ and $5$ as needed.  The relative numerical values of the
contributions to $c_7^{eff}(\mu)$ with $\mu=m_c=1.5$ GeV are
\beqa
\lefteqn{c_7^{eff}(m_c)  \simeq  \Re c_7^{eff}(m_c) \,,} \nonumber \\
& = & 0.458\, c_7(M_W) + 0.125\, c_8(M_W) - 0.312[c_{2a}(M_W)+c_{2b}(M_W)] \,,
\nonumber \\
& = & 0.458(-0.241\times 10^{-6}) + 0.125(-0.139\times 10^{-5}) -0.312
\left( {-\lambda_s-\lambda_d\over\lambda_b}\right)  \,,\nonumber \\
& = & 0.458(-0.241\times 10^{-6}) + 0.125(-0.139\times 10^{-5}) -0.312
 \,,
\label{cnumb}
\eeqa
where the CKM unitarity condition $\lambda_s + \lambda_d = - \lambda_b$
has been used to simplify the final term.
Incidentally, it should be stressed that the choice of $-\lambda_b$
as a prefactor in Eq.~(\ref{cheff}) was quite arbitrary, and we could have
pulled out some other factor, say $\lambda_s$ (or $\lambda_d$).
This would have affected the Wilson coefficients at the matching scale,
but the final result would have remained, as it must, unchanged.

We now compute the branching fraction. We evaluate $\alpha_s$
in the $\overline {MS}$ scheme (using $\alpha_s(M_Z)=0.124$ as before) and
extend the range down to the charm scale using the Bernreuther matching
conditions$^{\cite{bern}}$ at the threshold $\mu=m_b=4.87$ GeV.  Note that
we have also taken the CKM matrix elements to be real and have neglected
any possible imaginary components.  Given the small
values of $c_{7,8}(M_W)$, this approximation is well justified.
It is clear from the above that the $c_2(M_W)$ term completely dominates,
due to the small contributions to $c_{7,8}(M_W)$ from the light internal
quark masses.  This is in stark contrast to $b\to s$ transitions (and
likewise to $s\to d$), where the heavy internal $t$-quark forces the
the magnetic dipole coefficients to be competitive with $c_2(M_W)$.  This
can be seen explicitly by comparing the above with Eq.~(\ref{c7eff}).
The QCD-corrected branching fraction is then
\begin{eqnarray}
{\rm B}_{c\to u\gamma} & = &  {6\alpha\over \pi}\cdot
\left| {V^*_{cb}V_{ub}\over V_{cs}}\right|^2 \cdot {|c_7^{eff}(\mu)|^2\over
g(m_s/m_c)+{ \mbox{$|V_{cd}|^2$}\over \mbox{$|V_{cs}|^2$}}
g(m_d/m_c) } \cdot {\rm B}_{D^+\to X\ell^+\nu} \,, \nonumber \\
& = & \left( 4.21 - 7.94 \right) \times 10^{-12} \,,
\label{qcdchbr}
\end{eqnarray}
where the lower (upper) value in the numerical range corresponds to the scale
$\mu=2m_c (m_c)$.
We see that the effects of the QCD corrections are quite dramatic in
charm radiative decays, and that the rate is given almost entirely as a
consequence of operator mixing.  The stability of this result can be tested
once the complete next-to-leading order corrections to the magnetic dipole
transitions are known.

Finally, we wish to comment further on the CKM dependence of the
$|\Delta b| = 1$ and $|\Delta c| = 1$ effective hamiltonians.
We consider each case separately:

{\it $\ (i)\ |\Delta b| = 1$ transition}: Here, the the $t$-quark
contribution is seen to dominate in every respect.  Thus, for the
dipole operators $O_{7}$ and $O_8$, the $u$-quark and $c$-quark
loops are omitted because they are numerically tiny ({\it e.g.}, see Table~2).
Likewise, due to the smallness of the $u$-quark CKM factors,
the approximation is made in the literature$^{\cite{gsw}}$ to omit any
current-current operators containing $u$-quark fields.  This explains why
only the $c$-quark dependent operators $O_{1,2}$ appear in the
$|\Delta b| = 1$ operator basis of Eq.~(\ref{effh1}).  This assumption
also explains another aspect of the analysis.  Ordinarily one would
expect $O_{1,2}$ to be accompanied by the CKM factor $-\lambda_c$,
yet it is the prefactor $\lambda_t$ which appears in the effective
hamiltonian of Eq.~(\ref{effh}).  This is because the tiny value of
$\lambda_u$ has allowed one to write the CKM unitarity relation as
$\lambda_c \simeq - \lambda_t$ and thus remove dependence upon $\lambda_u$.

{\it $(ii)\ |\Delta c| = 1$ transition}: In this case, the CKM
dependence is more complicated since no single quark-loop is dominant.
One must expand the operator basis as we did in Eq.~(\ref{newop}).  However,
we wish to take note of a seemingly remarkable feature which occurs
upon carrying out the RG analysis.
The operators $O_{2a}$ and $O_{2b}$ turn out to have
equal anomalous dimensions and thus $c_{2a}$ and $c_{2b}$ have the
{\it same} numerical coefficient in Eq.~(\ref{cnumb}).   The most
elegant way to understand this result is to exploit the $U$-spin
symmetry present in the system of operators $O_{1a,1b}$ and $O_{2a,2b}$.
Thus, suppose instead of proceeding as we did, we
replaced the operators $O_{2a}$ and $O_{2b}$ of Eq.~(\ref{newop}) with
the equivalent set $O_{2a}'$ and $O_{2b}'$, where
\beqa
O_{2a} \to  O_{2a}' &\equiv& {1\over 2}(O_{2a} + O_{2b}) \nonumber \\
&=& {1\over 2}\left[ (\bar u_\alpha\gamma_\mu P_Ls_\alpha)(\bar
s_\beta\gamma^\mu P_Lc_\beta) + (\bar u_\alpha\gamma_\mu P_Ld_\alpha)
(\bar d_\beta\gamma^\mu P_Lc_\beta)\right] \,, \nonumber \\
O_{2b} \to  O_{2b}' &\equiv& {1\over 2}(O_{2a} - O_{2b}) \label{sp1} \\
&=& {1\over 2}\left[(\bar u_\alpha\gamma_\mu P_Ls_\alpha)
(\bar s_\beta\gamma^\mu
P_Lc_\beta) - (\bar u_\alpha\gamma_\mu P_Ld_\alpha)
(\bar d_\beta\gamma^\mu P_Lc_\beta) \right]\ , \nonumber
\eeqa
along with a corresponding replacement of coefficients,
\beq
c_{2a} \to c_{2a}' \qquad {\rm and} \qquad c_{2b} \to c_{2b}' \ \ .
\label{sp2}
\eeq
The matching conditions for the modified coefficients would then have become
\beq
c_{2a}' (M_W) = {\lambda_s + \lambda_d \over -\lambda_b} = 1
\qquad {\rm and} \qquad
c_{2b}' (M_W) = {\lambda_s - \lambda_d \over -\lambda_b} \ \ ,
\label{sp3}
\eeq
with analogous replacements made also for $O_{1a}$ and $O_{1b}$.
Since the mixing of $O_{2a}$ and $O_{2b}$ with $O_7$ has no dependence
on the mass of the internal $s$ and $d$ quarks, the operator $O_{2b}'$
does not contribute to the process $c \ra  u+ \gamma$ due to cancellation
between the $s$-quark and $d$-quark contributions.  This cancellation is
in fact just the manifestation of an underlying $U$-spin symmetry.  That is,
in the limit of neglecting quark mass, $O_{2b}'$ carries $U$-spin $1$
and thus cannot couple to a photon.  This decoupling occurs via the
very $s$-quark and $d$-quark cancellation under discussion.  As a consequence,
the other operator $O_{2a}'$ must have the the same anomalous dimension as
$O_{2a}$ under RG flow, and we obtain the result cited above.

As a corollary, it is clear that in the $|\Delta b| = 1$ transition, the
approximation made of omitting the $u$-quark current-current operators
is quite unnecessary.  One could just as easily deal with an expanded
operator basis containing $u$-quark fields analogous to that of
Eq.~(\ref{newop}) or by invoking an $SU(4)$ version
of $u$-quark/$c$-quark $U$-spin symmetry and proceeding as above.

\section{\bf Long Distance Pole Contributions}
As seen in the previous section, the short distance contributions to
charm radiative decays give very small branching ratios, even when
the large QCD enhancement is taken into account. There will, however, be
additional contributions that appear as exclusive modes for which the
momentum scale of the intermediate quarks is a strong interaction scale
and not the short distance scale $M_W$. This forces us to view the
intermediate states as hadrons rather than quarks. These long distance
contributions can be partitioned into two basic classes.  The first
corresponds, at the quark level, to annihilation diagrams
$c\bar{q}_1\to~q_2\bar{q}_3$ where a photon line is attached to any of
the four quark lines. In terms of hadronic degrees of freedom,
these give rise to the set of contributions which include the {\it pole
diagrams}.  The second type of contribution corresponds to the underlying
quark processes $c\to~q_1\bar{q}_2q$, followed by $\bar{q}_2q\to~\gamma$.
At the hadronic level, this is the so-called {\it vector meson dominance}
mechanism. We shall discuss the pole amplitudes in this
section, leaving consideration of the VMD mechanism for Section~4.

The pole amplitudes are but a subset of an entire class of long-distance
contributions, including the two-particle intermediate states and
proceeding to all higher $n$-particle intermediate states.  However,
of these the most phenomenologically accessible are the single-particle
or pole terms.  The relevant diagrams, appearing in Figs.~3(a,b), are seen
to fall into either of two basic classes.  We shall refer to transitions as
type~I if weak-mixing occurs before photon emission, {\it i.e.} if the
incoming $D$ meson experiences weak-mixing, and as type~II if photon emission
occurs before weak-mixing, {\it i.e.} if the final state meson is created
via weak-mixing.  In principle, the intermediate states occurring
in the type~I and type~II amplitudes consist respectively of all
possible virtual spin-zero and spin-one particles.  We shall find it
practicable, however, to take into account only the lightest such virtual
particles.
\phantom{xxxx}\vspace{0.03in}
\begin{center}
\begin{tabular}{c}\phantom{xxxxxxxxxxxxxxxxxxxxxx} \\
\phantom{xxxxxxxxxxxxxxxxxxxxxx} \\ \hline
\phantom{xxxxxxxxxxxxxxxxxxxxxx} \\
\phantom{xxxxxxxxxxxxxxxxxxxxxx} \\
\phantom{xxxxxxxxxxxxxxxxxxxxxx} \\
\phantom{xxxxxxxxxxxxxxxxxxxxxx} \\ \hline
\phantom{xxxxxxxxxxxxxxxxxxxxxx} \\
\phantom{xxxxxxxxxxxxxxxxxxxxxx} \\
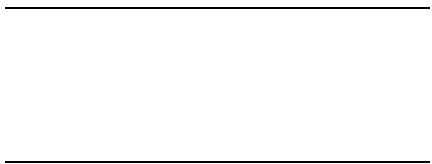
{Figure 3. Pole Contributions} \\
\end{tabular}
\end{center}
\vspace{0.08in}

In analyzing long range effects for flavor-changing $D$ decays, we shall
employ the effective weak hamiltonian of Bauer, Stech and
Wirbel$^{\cite{bsw}}$ (BSW),
\begin{equation}
{\cal H}_{\rm w} = -{G_{\rm F} \over \sqrt{2}}
\left[ :a_1 ({\bar u} d')({\bar s}' c) +
a_2 ({\bar s}' d')({\bar u} c) : \right] \ \ ,
\label{py1}
\end{equation}
where the colons denote normal-ordering and $d'$, $s'$ are the
CKM-mixed fields
\beqa
d' &=& V_{ud}~d + V_{us}~s  \ , \nonumber \\
s' &=& V_{cs}~s + V_{cd}~d  \ .
\label{py2}
\eeqa
We shall work in the $2\times 2$ basis of quark flavors,
\beq
{\bf V} = \left( \begin{array}{cc}
 V_{ud} & V_{us} \\
V_{cd} & V_{cs} \\  \end{array} \right) \simeq
\left( \begin{array}{cc}
 0.975 & 0.222 \\
-0.222 & 0.975 \\  \end{array} \right) \ \ .
\label{py3}
\eeq
Specific forms for the Cabibbo-favored and Cabibbo-suppressed
hamiltonians will be given shortly.  The quark fields
occur in left-handed combinations, denoted by
\beq
({\bar q}_1 q_2) \equiv {\bar q}_1 \gamma_\mu (1 + \gamma_5) q_2 \ \ ,
\label{py4}
\eeq
and $a_1, a_2$ are free parameters whose values will generally depend
on the mass scale being probed.  Here, they are determined by fitting to
$D\to {\bar K}\pi$ data$^{\cite{wir}}$,
\beq
a_1 (m_c^2) = 1.2 \pm 0.1 \ , \qquad a_2 (m_c^2) = -0.5 \pm 0.1  \ \ .
\label{py5}
\eeq

\begin{center}
{\bf Pole Amplitudes of Type~I}
\end{center}

Among the possible exclusive $D$ decays, the most promising for
experimental detection occur in the class of vector meson--photon
($V\gamma$) final states,
\beq
D({\bf p}) \to V ({\bf k}, \lambda ) + \gamma ({\bf q}, \sigma ) \ \ .
\label{py6}
\eeq
For these, the transition amplitude has the gauge invariant form
\beq
{\cal M}_{D\to V \gamma}  = \epsilon_\mu^\dagger
(k ,\lambda ) \epsilon_\nu^{\dagger} (q, \sigma)~
\left[ {\cal A}^{\rm pv} \left( p^\mu p^\nu - g^{\mu\nu} q
\cdot p \right) + i{\cal A}^{\rm pc} \epsilon^{\mu\nu\alpha\beta}
k_{\alpha} p_\beta \right] \ .
\label{py7}
\eeq
The parity-violating and parity-conserving amplitudes are denoted by
${\cal A}^{\rm pv}$ and ${\cal A}^{\rm pc}$ respectively, and each
carries the dimension of inverse energy.   Both amplitudes are generally
required because the weak interaction does not respect parity invariance.
The $D\to V \gamma$ decay rate is given by
\beq
\Gamma_{D \to V \gamma} = {|{\bf q}|^3 \over 4\pi} ~
(|{\cal A}^{\rm pv}|^2 + |{\cal A}^{\rm pc}|^2 ) \ \ ,
\label{py8}
\eeq
where ${\bf q}$ is the decay momentum in the $D$ rest frame,
\beq
|{\bf q}| = {m_{\rm D}^2 - m_V^2 \over 2 m_{\rm D}} \ \ .
\label{py10}
\eeq
Which particular combination of the parity-conserving and
parity-violating amplitudes contributes to the decay process will
depend upon the weak-mixing amplitude.  In principle,
a charm meson can mix with a sequence of either scalar $\{ S_n \}$ or
pseudoscalar $\{ P_n \}$ mesons.  Although some work on
scalar-mixing has been done,$^{\cite{eg}}$ the outcome
is rather model-dependent because detailed experimental and
theoretical understanding about scalar states is lacking.
In this paper, we shall therefore consider only the weak-mixing of
charm mesons with light pseudoscalar mesons
and thus work with only parity-conserving (PC) pole amplitudes.

It is appropriate at this point to comment on the notation
to be employed from this point on in both Section~3 and Section~4.
We shall denote $f_P$ as the decay
constant of pseudoscalar meson $P$ and define $h_{V\gamma P}$ as the
coupling constant for the EM interaction vertex of the photon $\gamma$ with
the mesons $V$,~$P$.  Also, the decay constant of vector meson $V$ is given
in terms of the $V$-to-vacuum matrix element of the vector current,
\beq
\langle 0 | V_\mu^a | V^b ({\bf q},\lambda ) \rangle = \delta^{ab}
{m_V^2 \over f_V} \epsilon_\mu^* ({\bf q},\lambda )
\equiv \delta^{ab} g_V \epsilon_\mu^* ({\bf q},\lambda ) \ \ .
\label{vec}
\eeq
Note that we define two equivalent parameterizations, $g_V$ (with units
of GeV$^2$) and $f_V$ (dimensionless), for the vector decay constant.
We have found that employing $g_V$ in the discussion of pole amplitudes
alleviates notational confusion which would otherwise occur between the
vector and pseudoscalar decay constants $f_V$ and $f_P$.  However, it is
traditional to use $f_V$ in discussing VMD amplitudes, and we do so in
Section~4.  The constants $\{ f_V\}$ are obtained from $\Gamma_{V\to
l^+l^-}$ data and have recently been compiled in Table~1 of Ref.~\cite{gp}.

Now, a pseudoscalar state $P$ which is created by weak mixing will propagate
virtually until it eventually decays into the final state.  This latter
transition is electromagnetic and hence parity-conserving.  It has
the amplitude
\beq
{\cal M}_{V \gamma P}  =  h_{V\gamma P}\epsilon_\mu^\dagger
(k ,\lambda ) \epsilon_\nu^{\dagger} (q, \sigma)
\epsilon^{\mu\nu\alpha\beta} k_{\alpha} p_\beta \ \ .
\label{py11}
\eeq
The absolute value of the coupling constant $h_{V\gamma P}$ can be inferred
phenomenologically by using
\beq
|h_{V\gamma P}|^2 = \left\{ \begin{array}{ll}
12\pi \Gamma_{V \to P \gamma}/ |{\bf q}|^3 &
(M_V > M_P) \\
4\pi \Gamma_{P \to V \gamma}/ |{\bf q}|^3  &
(M_P > M_V)\ . \end{array}\right.
\label{py13}
\eeq
The general type~I decay amplitude ${\cal A}_I$ for $D\to V\gamma$
is then given by
\beq
{\cal A}_I^{\rm pc}(D\to V\gamma) = \sum_n \ h_{V\gamma P_n} \cdot
{1\over m_D^2 - m_{P_n}^2}\cdot\langle P_n | {\cal H}^{\rm (eff)}_{\rm w}
| D \rangle \ \ .
\label{py14}
\eeq
With Fig.~3(a) as a guide, the notation should be self-evident.

Predictions for $D\to V\gamma$ decay amplitudes will be obtained below
in terms of both type I and type II pole amplitudes, and in the next
section we shall do the same by using VMD amplitudes.  We can, however,
accomplish somewhat more.  In principle,
the discussion for $V\gamma$ final states extends to a larger set
of meson-photon final states $M\gamma$, where the only restriction on
meson $M$ is that it have spin greater than zero.  For each different
type of $M\gamma$ final state, there will be a gauge invariant
$D$-decay amplitude like Eq.~(\ref{py7}) and an $M\gamma P$ interaction
vertex like Eq.~(\ref{py11}).  However, the generic form of Eq.~(\ref{py14})
continues to hold, except that $h_{V\gamma P_n}$ is replaced by
$h_{M\gamma P_n}$.  Of course, to have predictive power
requires knowledge of the $h_{M\gamma P_n}$ coupling constant.
Fortunately, much has been learned about radiative decays in light
meson systems over the years.  In particular, there are varying amounts of
experimental evidence for $17$ such transitions in the listing
of Ref.~\cite{pdg}.
Of these, $10$ involve $1^- \to 0^-$ mesonic transitions, $3$ involve
$2^+ \to 0^-$, $2$ involve $1^+ \to 0^-$ and $2$ involve $0^- \to 1^-$.
This information allows us to extend the analysis of type I
amplitudes from just $V\gamma$ final states to include both $A\gamma$
and $T\gamma$ configurations as well, where `$A$' and `$T$' stand for
axialvector and tensor mesons respectively.  The $A\gamma$ final states
are very analogous to the $V\gamma$ decays in that the coupling constant
$h_{A\gamma P_n}$ is found via Eq.~(\ref{py13}) and the $D\to A\gamma$
decay amplitude has the same form as Eq.~(\ref{py8}).  For the $T\gamma$
final states, one uses instead
\beq
|h_{T\gamma P}|^2 =
{40\pi \Gamma_{T \to P \gamma}\over |{\bf q}|^5}
\label{p16}
\eeq
as well as
\beq
\Gamma_{D \to T \gamma} = {|{\bf q}|^5 \over 4\pi}
{}~|{\cal A}_I^{\rm pc}|^2  \ \ .
\label{py8a}
\eeq
To summarize, we shall include in our study of type I amplitudes
certain $D\to A\gamma$ and $D\to T\gamma$ transitions.  For the
type II or VMD amplitudes, however, we shall limit our calculations
to just the $V\gamma$ final states.

\vspace{0.3in}

\noindent {\it Cabibbo-favored (CF) transitions}:

In this case, the BSW hamiltonian becomes
\begin{equation}
{\cal H}^{\rm (CF)}_{\rm w} = - V_{ud} {V_{cs}}^*{G_{\rm F} \over \sqrt{2}}
\left[ :a_1 ({\bar u} d)({\bar s} c) +
a_2 ({\bar s} d)({\bar u} c) : \right] \ \ .
\label{p7}
\end{equation}
The calculation of weak-mixing matrix elements of $D$'s with the light
pseudoscalar mesons is straightforward and results are tabulated in
Table~4.  The fact that these mixing amplitudes are evaluated in vacuum
saturation makes the forms in Table~4 easy to interpret.  Thus, for example,
in Cabibbo-favored $D_s^+$ decay, it is the term in the BSW hamiltonian
with coefficient $a_1$ which contributes, and as such, the weak-mixing
matrix element is naturally proportional to the decay constants $f_\pi$
and $f_{D_s^+}$.
\phantom{xxxx}\vspace{0.1in}
\begin{center}
\begin{tabular}{lc}
\multicolumn{2}{c}{Table~4 {Cabibbo-favored Mixing Amplitudes}}\\ \hline\hline
Mixing & Matrix Element \\ \hline
$D_s^+ \to \pi^+$ & $a_1 V_{ud} {V_{cs}}^* f_\pi f_{D_s^+}
m_{D_s^+}^2 G_F/\sqrt{2}$ \\
$D^0 \to {\bar K}^0$ & $a_2 V_{ud} {V_{cs}}^* f_K f_{D} m_{D^0}^2
G_F/\sqrt{2}$ \\
$D^+ \to \pi^+$ & $0$ \\ \hline\hline
\end{tabular}
\end{center}
\phantom{xxxx}\vspace{0.1in}
For the decay constants of the light mesons we use
\beq
f_\pi = 131~{\rm MeV} \qquad {\rm and} \qquad f_K = 161~{\rm MeV} \ \ .
\label{p8}
\eeq
The present situation for the decay constants $f_D$ and $f_{D_s}$
of the charm mesons is somewhat problematic.  Experiment provides
the upper limit for $f_D$,
\begin{equation}
f_D < 290~{\rm MeV} \ \ ,
\label{p9}
\end{equation}
as obtained from the branching ratio determination
$Br_{D^+ \to \mu^+ \nu_\mu} <  7.2 \times 10^{-4}$ at $90 \%$
confidence-level$^{\cite{mark3}}$.  Thus only theoretical
estimates exist for $f_D$.  These occur in three categories,
lattice theoretic$^{\cite{cb}}$, QCD sum rule$^{\cite{dom}}$ and
quark model fits to color-hyperfine mass splittings$^{\cite{ros}}$.
Estimates fall in the range $185 < f_D ({\rm MeV}) < 262$.  We
shall adopt the value
\beq
f_D^{\rm latt.} \simeq 216~{\rm MeV} \ \ ,
\label{p10}
\eeq
which is an average over the lattice estimates$^{\cite{shig}}$
and falls between the other two types of determinations.

Recently, the following experimental results (in units of MeV)
for $f_{D_s}$ were announced by CLEO$^{\cite{cleo6}}$,
WA75$^{\cite{wa75}}$ and BES$^{\cite{bes}}$,
\beq
f_{D_s} = \left\{ \begin{array}{ll}
344 \pm 37 \pm 67 & ({\rm CLEO})\\
232 \pm 45 \pm 52 & ({\rm WA75}) \\
434^{+153~+35}_{-133~-33} & ({\rm BES}) \ \ , \end{array}\right.
\label{p11}
\eeq
where the CLEO value is inferred from the ratio $\Gamma_{D_s^+ \to \mu^+
\nu_\mu}/ \Gamma_{D_s^+ \to \phi \pi^+}$ along with the branching ratio
$B_{D_s^+ \to \phi \pi^+}$.  In our numerical analysis, we shall use the
following weighted average of the above decay constants,
\beq
f_{D_s}^{\rm expt.} = 299\ {\rm MeV} \ \ .
\label{p13}
\eeq
For the sake of comparison, we note that this value is somewhat
larger than the central value of a weighted average taken from
a compilation of existing lattice estimates,$^{\cite{shig}}$
\beq
f_{D_s}^{\rm latt.} = 242\ {\rm MeV} \ \ .
\label{p14}
\eeq
The only other ingredients needed are the radiative coupling
constants $h_{M\gamma P}$, which were defined earlier.
Putting together all the necessary ingredients and ranging over
the set of final state mesons $M = \rho(770)$, $K^*(892)$, $b_1 (1235)$,
$a_1 (1270)$, $a_2 (1320)$ and $K_2^*(1430)$ yields the magnitudes
of type~I pole-model amplitudes (in units of GeV$^{-1}$) given in Table~5.
\phantom{xxxx}\vspace{0.1in}
\begin{center}
\begin{tabular}{lc}
\multicolumn{2}{c}{Table~5 {Type~I Cabibbo-favored Decay}}\\ \hline\hline
Mode & $|{\cal A}_I^{\rm pc}|~({\rm GeV}^{-1})$ \\ \hline
$D_s^+ \to \rho^+\gamma$ & $8.2\times 10^{-8}$ \\
$D_s^+ \to b_1^+ (1230)\gamma$ & $7.2\times 10^{-8}$ \\
$D_s^+ \to a_1^+ (1270)\gamma$ & $1.2\times 10^{-7}$ \\
$D_s^+ \to a_2^+ (1320)\gamma$ & $2.1\times 10^{-7}$ \\
$D^0 \to {\bar K}^{*0}\gamma$ & $5.6\times 10^{-8}$ \\ \hline\hline
\end{tabular}
\end{center}
\phantom{xxxx}\vspace{0.1in}
These values should be considered as upper bounds for the
following reason.  We have considered the lightest possible intermediate
states, pions and kaons, because only for these particles is there
sufficient data for determining coupling constants.  However,
the pion and kaon intermediate states propagate far off-shell.
Instead of having a squared momentum near the mass-shell value
$q^2 = m_\pi^2$, the virtual pion must carry $q^2 = m_{D_s}^2 \gg
m_\pi^2$ and similarly for the kaon.  This effect could well suppress
the transition amplitude.

In principle, one is to sum over all pion-like and kaon-like
intermediate states.  Other possible contributions should be
heavier and thus less affected by this suppression effect.  For
pion-like intermediate states, the next state in order of increasing mass
would be $\pi (1300)$ and beyond that the unconfirmed state $\pi (1770)$.
Although there is not sufficient data to make a numerical estimate of
their effect, we can anticipate for such states that

\ \ (i) the propagator contribution will indeed be larger,

\ (ii) the weak-mixing between a ground state $D$ meson and a
radially excited meson $P_n$ will be wave-function suppressed,

(iii) the radiative coupling constant $h_{M \gamma P_n}$ might well be
relatively smaller due to phase space competition with other decay
modes of the massive meson $P_n$.

We would expect the net result of these effects to decrease the overall
contribution from the excited states.
\vspace{0.3in}

\noindent {\it Cabibbo-suppressed (CS) transitions}:

The weak-mixing now proceeds according to the weak hamiltonian
\beqa
{\cal H}^{\rm (CS)}_{\rm w} &=& - {G_{\rm F} \over \sqrt{2}}
\left[ : a_1 \left(V_{ud} {V_{cd}}^*({\bar u} d)({\bar d} c) +
V_{us} {V_{cs}}^*({\bar u} s)({\bar s} c)\right)\right.
\nonumber \\
& & \left. +
a_2 \left( V_{us} {V_{cs}}^*({\bar s} s)({\bar u} c) +
V_{ud} {V_{cd}}^*({\bar d} d)({\bar u} c)\right) : \right] \ \ ,
\label{p17}
\eeqa
The action of the $({\bar d} d)$ and $({\bar s} s)$ operators
on the vacuum when expressed in terms of the pseudoscalar meson
states becomes
\beqa
({\bar d} d) &=& -0.7071 \pi^0 + 0.58 \eta + 0.40 \eta' \nonumber \\
({\bar s} s) &=& -0.57 \eta + 0.82 \eta' \ \ ,
\label{p18}
\eeqa
where an $\eta - \eta'$ mixing angle $\theta_{\rm P} = -20^o$
is adopted.$^{\cite{dgh}}$  In addition, we take$^{\cite{dgh}}$
\beq
f_\eta \simeq f_{\eta'} \simeq f_\pi \ \ .
\label{eta}
\eeq
The mixing amplitudes which are relevant for Cabibbo-suppressed decay
appear in Table~6.  Observe that we have simplified the notation
for $D^0$ transitions by expressing $V_{ud}{V_{cd}}^*$ in terms of
$V_{us}{V_{cs}}^*$.
\phantom{xxxx}\vspace{0.1in}
\begin{center}
\begin{tabular}{lc}
\multicolumn{2}{c}{Table~6 {Cabibbo-suppressed Mixing Amplitudes}}\\
 \hline\hline
Mixing & Matrix Element \\ \hline
$D^+ \to \pi^+$ & $a_1 V_{ud} V_{cd}^* f_\pi f_{D^+} m_{D^+}^2 G_F/\sqrt{2}$ \\
$D_s^+ \to K^+$ & $a_1 V_{us} V_{cs}^* f_K f_{D_s} m_{D_s}^2 G_F/\sqrt{2}$ \\
$D^0 \to \pi^0$ & $0.7071 a_2 V_{us} V^*_{cs} f_{D^0} f_{\pi} m_{D^0}^2
 G_F/2$ \\
$D^0 \to \eta$ & -$1.15 a_2 V_{us} V^*_{cs} f_{D^0} f_{\eta} m_{D^0}^2
 G_F/2$ \\
$D^0 \to \eta'$ & $0.42 a_2 V_{us} V^*_{cs} f_{D^0} f_{\eta'}
m_{D^0}^2 G_F/2$ \\ \hline\hline
\end{tabular}
\end{center}
\phantom{xxxx}\vspace{0.1in}

The analysis for Cabibbo-suppressed decays proceeds analogous to
that for Cabibbo-favored decays, with one significant complication.
For each of the Cabibbo-favored transitions, only one amplitude
contributes.  For $D^0$ decay, however, all the Cabibbo-suppressed pole
amplitudes contain a sum over $\pi^0$, $\eta$ and $\eta'$
intermediate states.  It is important to get the relative phases
of the interfering amplitudes correct.  We have therefore
performed an analysis of the nine $V^0 \to P^0 \gamma$ couplings
in light of the most recent data, where $V^0 = \rho^0$, $\omega^0$,
$\phi^0$ and $P^0 = \pi^0$, $\eta$ and $\eta'$.$^{\cite{kjm}}$
The magnitudes of the Cabibbo-suppressed amplitudes are displayed in
Table~7.
\phantom{xxxx}\vspace{0.1in}
\begin{center}
\begin{tabular}{lc}
\multicolumn{2}{c}{Table~7 {Type~I Cabibbo-suppressed Decay}}\\
\hline\hline
Mode & $|{\cal A}_I^{\rm pc}|~({\rm GeV}^{-1})$ \\ \hline
$D^+ \to \rho^+\gamma$ & $1.3\times 10^{-8}$ \\
$D^+ \to b_1^+(1230)\gamma$ & $1.2\times 10^{-8}$ \\
$D^+ \to a_1^+(1270)\gamma$ & $4.9\times 10^{-9}$ \\
$D^+ \to a_2^+(1320)\gamma$ & $3.4\times 10^{-8}$ \\
$D_s^+ \to K^{*+}\gamma$ & $2.8\times 10^{-8}$ \\
$D_s^+ \to K_2^{*+}(1430)\gamma$ & $6.0\times 10^{-8}$ \\
$D^0 \to \rho^0\gamma$ & $4.8\times 10^{-9}$ \\
$D^0 \to \omega^0\gamma$ & $6.1\times 10^{-9}$  \\
$D^0 \to \phi^0\gamma$ & $7.4\times 10^{-9}$  \\ \hline\hline
\end{tabular}
\end{center}
\phantom{xxxx}\vspace{0.1in}

\begin{center}
{\bf Pole Amplitudes of Type~II}
\end{center}

Analogous to the type~I $D\to V\gamma$ decay amplitude of Eq.~(\ref{py14})
we have
\beq
{\cal A}_{II}^{\rm pc}(D\to V\gamma) = \sum_n \ \langle V |
{\cal H}^{\rm (eff)}_{\rm w} | D^*_n \rangle \cdot
{1\over m_D^2 - m_{D^*_n}^2}\cdot h_{D^*_n \gamma D}
\label{tpii}
\eeq
for the corresponding type~II transition.  From the viewpoint of
phenomenology, the type~II transitions are more problematic than are
those of type~I because less experimental input is available.  Thus,
we shall need to rely a bit more heavily on theoretical predictions.

The first difficulty is that the couplings $h_{D^{*0}\gamma D^0}$,
$h_{D^{*+}\gamma D^+}$, and $h_{D_s^{*+}\gamma D_s^+}$ have not
yet been experimentally measured.  This is because, although the relevant
photonic branching ratios have been measured, only upper bounds exist
for the full widths of the associated spin-one exited states, $D^{*0}$,
$D^{*+}$ and $D_s^{*+}$,
\beqa
\Gamma_{D^{*0}} < 2100~{\rm keV} \quad &\Longrightarrow& \quad
\Gamma_{D^{*0}\to D^0\gamma} < 764~{\rm keV} \nonumber \\
\Gamma_{D^{*+}} < 131~{\rm keV} \quad &\Longrightarrow& \quad
\Gamma_{D^{*+}\to D^+\gamma} < 1.44~{\rm keV} \label{p20} \\
\Gamma_{D_s^{*+}} < 4500~{\rm keV} \quad &\Longrightarrow& \quad
\Gamma_{D_s^{*+}\to D_s^+\gamma} < 4500~{\rm keV} \ . \nonumber
\eeqa
Fortunately, predictions for the $\Gamma_{D^{*0}\to D^0\gamma}$,
$\Gamma_{D^{*+}\to D^+\gamma}$ and $\Gamma_{D_s^{*+}\to D_s^+\gamma}$
transitions have appeared in the literature
recently.$^{\cite{{aip},{cdn},{ox},{fr}}}$  There is some
spread in predictions, and so we choose the representative values,
\beq
\Gamma_{D^{*0}\to D^0\gamma} = 20~{\rm keV}~, \quad
\Gamma_{D^{*+}\to D^+\gamma} = 0.5~{\rm keV} ~, \quad
\Gamma_{D_s^{*+}\to D_s^+\gamma} = 0.3~{\rm keV} \ ,
\label{p21}
\eeq
which implies
\beqa
h_{D^{*0}\gamma D^0} &=& 0.542~{\rm GeV}^{-1}\ , \nonumber \\
h_{D^{*+}\gamma D^+} &=& -0.087~{\rm GeV}^{-1} \ , \label{p21a} \\
h_{D_s^{*+}\gamma D_s^+} &=& -0.066~{\rm GeV}^{-1}  \nonumber \ \ ,
\eeqa
where we have adopted the phases implied by the quark model.
A rough check on whether the above values are reasonable is
afforded by the nonrelativistic quark model, in which
\beqa
h_{D^{*0}\gamma D^0} &=& 2e\left[ {1\over M_c} +  {1\over M_u} \right] \ ,
\nonumber \\
h_{D^{*+}\gamma D^+} &=& e\left[ {2\over M_c} - {1\over M_d}\right]\ ,
\label{p21b} \\
h_{D_s^{*+}\gamma D_s^+} &=& e\left[ {2\over M_c} - {1\over M_s}\right]\ ,
\nonumber
\eeqa
where the $\{ M_k \}$ are constituent quark masses, distinct from
the current masses $\{ m_k\}$ of Section~2.  If we take
$M_c \simeq 1.64$~GeV, as implied by a fit to $D$ and $D^*$ masses,
then the relations in Eqs.~(\ref{p21}-\ref{p21b})
yield $M_u \simeq M_d = 0.48$~GeV and $M_s \simeq 0.53$~GeV.

The other of the difficulties concerns the weak-mixing matrix elements.
For type~II transition amplitudes, the mixing occurs between
charm and light vector mesons, as in
\beq
\langle \rho^+ | {\cal H}^{\rm (eff)}_{\rm w} | D_s^{*+} \rangle \simeq
a_1 V_{ud} V_{cs} g_{\rho^+} g_{{\rm D}_s^{*+}} G_F/\sqrt{2} \ \ .
\label{p22}
\eeq
In the above, the $g_V$ are the vector meson decay constants defined in
Eq.~(\ref{vec}) and whose determination we shall discuss shortly.  As with
the type~I amplitudes, we have employed vacuum saturation.  To determine
the action of the $({\bar d} d)$ and $({\bar s} s)$ operators upon the
vacuum we employ the ideally-mixed vector meson states, so that
\beq
({\bar d} d) = {\omega - \rho^0 \over \sqrt{2}}
\qquad {\rm and} \qquad ({\bar s} s) = \phi \ \ .
\label{p22a}
\eeq
For the light $1^-$ mesons, the collection $\{ g_V \}$ of vector decay
constants can be determined by referring to the vacuum-to-meson matrix
elements of $J^\mu_{\rm em}$ given in Table~1 of Ref.~\cite{gp}.
Together with isospin and $SU(3)$ relations along with quark model
insights, these generate all the needed values, {\it e.g.}
\beq
g_{\rho^+} \simeq 0.17~{\rm GeV}^2 \ , \quad
g_{K^*} \simeq {m^2_{K^*} \over m^2_\rho} g_{\rho^+} \simeq 0.22~{\rm GeV}^2
\ , \quad \dots \ \ .
\label{p23}
\eeq
To estimate the $D_s^{*+}$ and $D^{*0}$ decay constants, we invoke
the heavy-quark-symmetry relations,
\beqa
g_{D_s^{*}} &=& m_{D_s} f_{D_s} \simeq 0.588~{\rm GeV}^2 \ , \nonumber \\
g_{D^{*}} &=& m_{D} f_{D} \simeq 0.403~{\rm GeV}^2 \ \ .
\label{p24}
\eeqa
The magnitudes of the type~II amplitudes thus calculated are given in
Table~8.
\phantom{xxxx}\vspace{0.1in}
\begin{center}
\begin{tabular}{lc}
\multicolumn{2}{c}{Table~8 {Type~II Decays}}\\ \hline\hline
Mode & $|{\cal A}_{II}^{\rm pc}|~({\rm GeV}^{-1})$ \\ \hline
$D_s^+ \to \rho^+\gamma$ & $1.9\times 10^{-8}$ \\
$D^0 \to {\bar K}^{*0}\gamma$ & $5.9\times 10^{-8}$ \\
$D^+ \to \rho^+\gamma$ & $3.6\times 10^{-9}$ \\
$D_s^+ \to K^{*+}\gamma$ & $5.1\times 10^{-9}$ \\
$D^0 \to \rho^0\gamma$ & $4.7\times 10^{-9}$ \\
$D^0 \to \omega^0\gamma$ & $6.9\times 10^{-9}$  \\
$D^0 \to \phi^0\gamma$ & $1.6\times 10^{-8}$  \\ \hline\hline
\end{tabular}
\end{center}
\phantom{xxxx}\vspace{0.1in}

$D^*$ excitations with spins not equal to one will not contribute to
type~II amplitudes if we adhere strictly to the hamiltonian of
Eq.~(\ref{py1})$^{\cite{gp}}$ and continue to work within the
vacuum saturation framework.  The reason is that mesons
with $J>1$ cannot have a nonzero matrix element with the vacuum
via the current ${\bar q}\gamma_\mu (1 + \gamma_5) c$.  The possibility
of an intermediate charm meson with $J=0$ is disallowed since
it could only mix with a final-state $J=0$ particle and the decay of a
spinless particle to another spinless particle plus a photon is
forbidden.

Although we have considered just final state
vector mesons in Table~8, it should be obvious that in principle
the spin-one intermediate $D^*$ states can also mix weakly with axialvector
final state mesons.  Unfortunately, one knows less about
the decay constants of axialvector mesons than one does of the
vector mesons.

\section{\bf Long Distance VMD Contributions}
The VMD contribution to charm meson radiative decay is depicted in
Fig.~4, where a $D$ meson is seen to (i) decay weakly into a final state
of a vector meson $V$ and a meson $M$ of nonzero spin, followed by
(ii) an electromagnetic VMD conversion of $V$ into a photon. Roughly
speaking, in the VMD approach the $D\to M\gamma$ amplitude is obtained
by multiplying the $D\to MV$ amplitude by the factor $e / f_V$ where $e$
is the electric charge and $f_V$ is the dimensionless version of the
vector meson decay constant defined in Eq.~(\ref{vec}).  It is important
to keep in mind that in the VMD process $D\to MV$, the vector meson $V$ is
off-shell.  Thus, to obtain the VMD amplitude for $D\to M\gamma$ will
require an extrapolation from $p_V^2=m_{V}^{2}$ to $p_V^2=0$ for both the
$V\to\gamma$ vertex and the $D\to MV$ transition.  For our considerations,
the main intermediate states will involve virtual rho and phi mesons.  We
shall employ the observation made in Ref.~{\cite{paul}} that the rho-gamma
vertex seems to be unaffected by the extrapolation whereas the phi-gamma
vertex is reduced by a factor of $\eta_\phi \simeq \sqrt{2}$.  In the
following, we will consider a number of examples for the case $M = V$, and
so we shall be working with VMD chains which begin with the process $D\to VV$.
Since the $VV$ final states have $L=0,1,2$ as allowed orbital angular
momentum values, the VMD amplitude will in general have a parity-conserving
part ${\cal A}_{\rm VMD}^{\rm pc}$ corresponding to the $VV$ $P$-wave and
a parity-violating part ${\cal A}_{\rm VMD}^{\rm pv}$ corresponding
to the $VV$ $S$-wave and/or $D$-wave.
\phantom{xxxx}\vspace{0.03in}
\begin{center}
\begin{tabular}{c}\phantom{xxxxxxxxxxxxxxxxxxxxxx} \\
\phantom{xxxxxxxxxxxxxxxxxxxxxx} \\ \hline
\phantom{xxxxxxxxxxxxxxxxxxxxxx} \\
\phantom{xxxxxxxxxxxxxxxxxxxxxx} \\
\phantom{xxxxxxxxxxxxxxxxxxxxxx} \\
\phantom{xxxxxxxxxxxxxxxxxxxxxx} \\ \hline
\phantom{xxxxxxxxxxxxxxxxxxxxxx} \\
\phantom{xxxxxxxxxxxxxxxxxxxxxx} \\
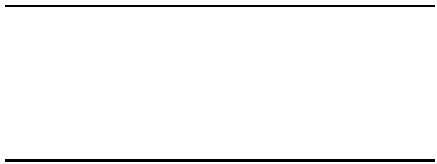
{Figure 4. VMD Contribution} \\
\end{tabular}
\end{center}
\vspace{0.08in}
In practice, there are two means for determining the $D\to MV$ part of a
VMD amplitude for $D\to M\gamma$:
\begin{enumerate}
\item One can input $D\to MV$ experimental data directly in order to
phenomenologically determine the $D\to MV$ amplitude.  In this approach,
it is crucial to maintain gauge invariance.  A careful discussion of how
to construct a gauge invariant amplitude was recently given in
Ref.~\cite{gp} (which considered this type of empirical VMD contribution
to $B\to K^*\gamma$), so we need not detail this procedure here.
Since the database for $D\to MV$ transitions is unfortunately small,
the ability to generate VMD amplitudes using this phenomenological
method is limited.
\item One can employ some theoretical description to model the $D\to MV$
amplitude.  Since the models currently available do not always reliably
reproduce branching ratios and polarizations
of final-state vector mesons in decays of heavy mesons,$^{\cite{dtp}}$
this method is also not beyond criticism.  For definiteness, we shall
continue to employ the BSW model$^{\cite{bsw}}$ introduced in
Section~3.$^{\cite{k}}$  Within this approach, the squared VMD amplitude
for the important case where $M$ is a vector meson becomes
\beqa
& & |{\cal A}_{\rm VMD}|^2=\frac{G_{F}^{2}|V_{cq}^{*}V_{qu}|^2}{2m_{D}^{2}
{\bf k}^2}a_{i}^{2}(m_{c}^{2})f^{2}_{X} I \nonumber \\
& & \times  \left[ (m_D+m_Y)^2A_{1}^{2}(q_{0}^{2}) +
\frac{4{\bf k}^2 m_{D}^{2}V^2(q_{0}^{2})
}{(m_D+m_Y)^2}\right] \times\left(\frac{4\pi\alpha}{f_{V}^{2}}\right) ~,
\label{bsw_rad}
\eeqa
where $|{\bf k}|$ is the photon spatial momentum, $q$ represents
either of the $d$ or $s$ light quarks, and $I$ is a process-dependent
isospin coefficient.  The BSW coefficients $a_{1}(m_{c}^{2})$ and
$a_{2}(m_{c}^{2})$ which correspond to the color-favored and
color-suppressed operators are given in Eq.~(\ref{py5}).
The remaining notation is explained by noting that in the factorization
approximation for $D\to MV$, one of the final state particles,
which we call $X$ (either~$M~{\rm or}~V$), couples directly to the vacuum
and the other, which we call $Y$ (either~$V~{\rm or}~M$), appears in the
$D$-to-$Y$ matrix element of the charged weak current $J^\mu_{\rm ch}$.
Thus the quantity $f_X$ is the decay constant of $X$, and
$A_1(q^2)$ and $V(q^2)$ are the semileptonic form factors defined by
\begin{eqnarray}
\lefteqn{\langle Y(p_Y)|J^\mu_{\rm ch}|D(P)\rangle =} \nonumber \\
& &\frac{2V(q^2)}{m_D+m_Y}\epsilon^{\mu\nu\rho\sigma}\epsilon^*_\nu
P_\rho  p_{Y\sigma} + 2m_Y iA_0(q^2)\frac{\epsilon^*.q}{q^2}q^\mu
\nonumber \\
& & +i\left[(m_D+m_Y)A_1(q^2) \epsilon^{*\mu} -
\frac{\epsilon^*.qA_2(q^2)}{m_D+m_Y}(P+p_y)^\mu \right. \nonumber \\
& &\left. - 2m_YA_3(q^2)\frac{\epsilon^*.q}{q^2}q^\mu\right] \ .
\label{ffs_def}
\end{eqnarray}
In the VMD amplitude, the form factors are to be evaluated at
$q_{0}^{2}=0$ if $X=V$ and at $q_{0}^{2}=m_{M}^2$ if $X=M$.
Throughout, we shall make use of the form factors as
measured$^{\cite{form}}$ in $D\to K^* l\nu$ and also employ $SU(3)$
relations as needed.  This should provide a good estimate of the
form factors appearing in the $D$-to-$\rho$ and $D$-to-$\phi$ matrix
elements.  Whenever the form factors are to be evaluated at momentum
transfers other than at $q^2 = 0$, we shall use a monopole form
to extrapolate from $q^2=0$. This amounts simply to dividing the
form factors at $q^2=0$ by the quantity $1-q^2/m_{\rm pole}^{2}$.
\end{enumerate}
In the following, we shall give VMD predictions for a number of
specific $D\to M\gamma$ decays, grouped as Cabibbo favored,
singly suppressed or doubly suppressed.  In the few cases where we can
employ both the above approaches, we shall refer to them respectively as
`Meth.~1' and `Meth.~2'.  Given the lack of abundant $D\to MV$ data,
however, we shall be forced to adopt the theoretical approach of Meth.~2 in
most cases.

Before we can proceed, there is another
topic which must be addressed, the dynamical complication of
significant Final State Interactions (FSI).  Although presumably not a
problem in $B\to MV$ decays, detectable FSI are known to exist
in the $D$-meson mass region.  This can produce an ambiguity in the VMD
analysis because FSI will inherently be part of any VMD amplitude
obtained from $D\to MV$ data, but will not be present in the BSW
construction.  It is difficult to remove the effect of FSI from
the phenomenological VMD amplitude because the vector meson $V$ is
to be taken off-shell, and FSI might have an important kinematic
dependence, {\it i.e.} the $p_{V}^{2}$ dependence of the FSI has also
to be taken into account.  Consequently, any FSI effects entering in
data may not be present to the same extent in the VMD amplitudes.
As regards the factorization construction (Meth.~2 above), the exclusion of
any FSI effects in the BSW amplitude amounts {\it de facto} to a
specific prescription for the $p_V^2$-dependence of the FSI.
There is some information on the $p_V^2$-dependence of the
$\gamma - V$ couplings and of certain matrix elements, but it
is not possible at this time to separate the two effects.  As we
show in the Appendix, the effect for $\rho$ emission in $A_2 (1320)$
decay can be as much as a factor of two.  By contrast, no such suppression
is seen in $\rho$-photoproduction, although in $\phi$-photoproduction
an effective reduction of about $\sqrt{2}$ in amplitude is observed and a
somewhat smaller effect of $\sqrt{1.5}$ is seen in $\omega$-photoproduction.
\vspace{0.1in}
%\vskip .1in
%\begin{abstract}
\begin{center}
{\bf Cabibbo-favored Modes}
\end{center}

\paragraph{$D^0\to {\bar K}{^{0*}}\gamma$:}
This is an instance in which the phenomenological approach is applicable
since experimental information on the $D\to MV$ intermediate state is
available.  There is a branching ratio determination$^{\cite{pdg}}$
$B_{D^0 \ra \bar{K}^{*0} \rho^0}=(1.6 \pm  0.4)\%$ and the amplitude is
known to be (i) almost all transverse and (ii) almost all
$S$-wave.\footnote{The Particle Data Group also lists branching ratios of
$(3.0 \pm 0.6) \%$ and $(2.1 \pm 0.6)\%$ for $S$-wave and D-wave
respectively.$^{\cite{pdg}}$  These values are completely consistent with
the fact that the total transverse mode (which must be entirely $S$-wave by
the absence of any $P$-wave) is $(1.6 \pm 0.5)\%$ and that the $S$-wave
(longitudinal) must cancel with the D-wave to produce the net zero
longitudinal branching ratio.} This allows us to write the VMD contribution
to $D^0 \ra \bar{K}^{*0} \gamma$ as
\begin{equation}
{\cal A}^{\rm pv}_{\rm VMD} = {e \over f_\rho} \cdot
\frac{a_{D^0\to \bar{K}^{*0} \rho^0}}{m_D E_\gamma}\ , \qquad \qquad
{\cal A}^{\rm pc}_{\rm VMD} \simeq 0 \ , \qquad ({\rm Meth.~1})
\end{equation}
where we follow the notation of Ref.~\cite{gp} and denote $a_{D^0\to
\bar{K}^{*0} \rho^0}$ as the phenomenological $S$-wave amplitude for
$D^0\to \bar{K}^{*0} \rho^0$.  With $\Gamma_{D^0 \to \bar{K}^{*0} \rho^0}
= 2.53\times 10^{-14}$~GeV and $a_{D^0\to \bar{K}^{*0} \rho^0} =
1.63\times 10^{-6}$~GeV, this yields ${\cal A}^{\rm pv}_{\rm VMD} (D^0
\ra \bar{K}^{*0} \gamma)$ of about $6.8\times 10^{-8}~{\rm GeV}^{-1}$.
The data on $D^0 \to {\bar K}^{*0}\rho^0$ is consistent with no
parity-conserving ($P$-wave) contribution.

Alternatively, the factorization approach of Eq.~(\ref{bsw_rad}) predicts
both amplitudes. In this case, we take $a_i=a_2$, and the vector meson
to be mixed with the photon is the $\rho^0$, so that $V=\rho^0$
and $X=\bar{K^{0*}}$.  The form factors needed are those entering
in $D\to\rho$ semileptonic transitions. Making use of the measured
$D\to K^*$ form factors implies $I=1/2$. To extrapolate the form factors
from $q^2=0$ to $q^2=m_{K^*}^{2}$, we use a monopole form where the
$D^*$ is the nearest singularity. The parity-violating and
parity-conserving amplitudes are
given in Eq.~(\ref{bsw_rad}) by the terms involving the $A_1$ and $V$
form factors respectively. Using $f_{K^*}=0.2~{\rm GeV}^2$ we obtain
\begin{equation}
{\cal A}_{\rm VMD}^{\rm pv}=5.1\times 10^{-8}~{\rm GeV}^{-1} \ , \qquad
{\cal A}_{\rm VMD}^{\rm pc} = 3.8\times 10^{-8}~{\rm GeV}^{-1} \ .
\qquad ({\rm Meth.~2})
\end{equation}
We notice that ${\cal A}_{\rm VMD}^{\rm pv}$ is in reasonable agreement with
the one obtained from the use of data from the nonleptonic mode, given the
large uncertainties involved in these predictions. Indeed, the
factorization estimate for the $D^0\to\bar{K^{0*}}\rho^0$ $S$-wave amplitude
gives $a_{D^0\to \bar{K}^{*0} \rho^0} =1.3\times 10^{-6}~{\rm GeV}^{-1}$
which is within 20\% of the experimental value. It also predicts a $P$-wave
branching fraction of $0.15\%$ for $D^0 \to \bar{K^{0*}}\rho^0$, which
is below the current upper limit of $0.30\%$.

\paragraph{$D^+_s\to\rho^+\gamma$: } The VMD amplitude for this decay
proceeds via $D^+_s\to\phi\rho^+$ followed by $\phi$-$\gamma$ mixing.
Although the branching ratio for $D^+_s \ra \phi \rho^+$ is known to be
($6.5 \pm_{1.8}^{1.6}\%$), no information on helicities or partial waves
exists, so we cannot apply the phenomenological method here.  Turning
instead to the factorization approach of Eq.~(\ref{bsw_rad}),
we have $X=V=\phi$ and $Y=M=\rho^+$. Therefore we require the
$D^+_s\to\phi$ semileptonic form factors evaluated at
$q_{0}^{2}=m_{\rho}^{2}$. Although there is experimental information
on these decays, the branching fraction and the form factors depend
strongly on $B_{D^+_s\to\phi\pi^+}$, which is still very uncertain.
Thus, again making use of $D\to K^*$ data, taking $I=1$ and with a
decay constant of $g_\rho\simeq 0.17~{\rm GeV}^2$, we find
\begin{equation}
{\cal A}^{\rm pv}_{\rm VMD}  =  3.2\times10^{-8}~{\rm GeV}^{-1}
\qquad {\rm and}
\qquad {\cal A}^{\rm pc}_{\rm VMD}  =  2.8\times10^{-8}~{\rm GeV}^{-1} \ \ .
\end{equation}
The Cabibbo-favored VMD amplitudes are summarized in Table~9.
\phantom{xxxx}\vspace{0.1in}
\begin{center}
\begin{tabular}{l|cc}
\multicolumn{3}{c}{Table~9 {Cabibbo-favored VMD Amplitudes}}\\
 \hline\hline
Mode & \multicolumn{2}{c}{$|{\cal A}_{\rm VMD}|~(10^{-8}~{\rm GeV}^{-1})$}
\\ \hline
& Parity-conserving & Parity-violating \\ \hline
$D^0 \to {\bar K}^{*0}\gamma$ & $3.8$ & 5.1-6.8 \\
$D_s^+ \to \rho^+\gamma$ & $3.2$ & $2.8$ \\ \hline\hline
\end{tabular}
\end{center}
\vspace{0.1in}
\begin{center}
{\bf Singly Cabibbo-suppressed Modes}
\end{center}

\paragraph{$D^0\to\rho^0\gamma$:}
This process can proceed via two different intermediate states, namely
$\rho^0\phi$ and $\rho^0\rho^0$. There is one known branching ratio
$B_{D^0 \ra \rho^0\phi}\ = (1.9 \pm 0.5)\times 10^{-3}$ with no
helicity (or partial wave) information.  Letting $\eta_T$ be the
transverse fraction of the observed branching ratio,
$\eta_S$ the $S$-wave fraction in the transverse mode, and $\eta_P$
the $P$-wave fraction, we then obtain for
the $S$-wave amplitude of $D^0 \ra \rho^0\phi$,
\begin{equation}
a_{D^0 \to \rho^0\phi} = m_D\sqrt{ 4\pi \Gamma_{D^0 \ra \rho^0\phi}
\eta_T \eta_S \over |{\bf k}|}
=  7\times 10^{-7} \sqrt{\eta_T \eta_S}~{\rm GeV}\ \ ,
\end{equation}
and for the corresponding $P$-wave,
\begin{equation}
{b_{D^0\to \rho^0 \phi}\over m_\phi m_\rho} =
\sqrt{4 \pi \Gamma_{D^0 \ra \rho^0\phi} \eta_T \eta_P \over
|{\bf k}|^3} = 1.46\times 10^{-6} \sqrt{\eta_T \eta_P}~{\rm GeV}^{-1}\ \ ,
\end{equation}
where again we employ the notation of Ref.~\cite{gp} in denoting
$b_{D^0\to \rho^0\phi}$ as the phenomenological $P$-wave
amplitude for $D^0\to \rho^0\phi$.  Then, multiplying by the
VMD factor $e/f_{\phi}$, we obtain the Meth.~1 estimate
\begin{equation}
{\cal A}^{\rm pv}_{\rm VMD}  =  0.60\times 10^{-8}~{\rm GeV}^{-1}
\qquad {\rm and} \qquad
{\cal A}^{\rm pc}_{\rm VMD}  =  1.0\times 10^{-8}~{\rm GeV}
\label{rhogam_1}
\end{equation}
for $\eta_T \sim 0.5$ and $\eta_S \sim 0.66$.  On the other hand, there is no
available experimental information for $D^0\to\rho^0\rho^0$, other than
$B_{D^0\to\pi^+\pi^-\pi^+\pi^-} = (8.3\pm0.9)\times 10^{-3}$ which can
be taken as an upper limit. Let us also estimate the $D^0 \to \rho^0 \gamma$
mode in the factorization approach, which can be used to predict both the
off-shell amplitudes, $D^0\to \rho^0\phi$ and $D^0\to \rho^0\rho^0$. In both
cases we need the $D^0\to\rho^0$ form factors, for which $I=1/2$. Using
Eq.~(\ref{bsw_rad}) we obtain
\begin{eqnarray}
{\cal A}_{\rm VMD}^{\rm pv}(\rho^0\phi \to \rho^0\gamma )&=&0.22\times
10^{-8}~{\rm Gev}^{-1}\nonumber\\
{\cal A}_{\rm VMD}^{\rm pc}(\rho^0\phi \to \rho^0\gamma )&=&0.18\times
10^{-8}~{\rm Gev}^{-1}\nonumber\\
{\cal A}_{\rm VMD}^{\rm pv}(\rho^0\rho^0 \to \rho^0\gamma )&=&0.75\times
10^{-8}~{\rm Gev}^{-1}\nonumber\\
{\cal A}_{\rm VMD}^{\rm pc}(\rho^0\rho^0 \to \rho^0\gamma )
&\simeq& 0 \ \ .
\label{rhogam_2}
\end{eqnarray}
Our estimate for the parity-conserving $\rho^0\rho^0 \to \rho^0\gamma$
transition is based on the observation that an on-shell $P$-wave $\rho^0\rho^0$
state is forbidden by Bose statistics and hence the associated off-shell
amplitude will be suppressed.  First, let us compare the first two rows
in Eq.~(\ref{rhogam_2}) with the results obtained in Eq.~(\ref{rhogam_1})
by making use of the $\rho^0\phi$ data.  We can see that the factorization
amplitudes are lower, as caused by smaller predictions for the
nonleptonic intermediate modes. In general, factorization predictions will
be modified by FSI.  For instance, in the case at hand there could be a large
enhancement due to $K^*\bar{K}^*\to\rho^0\phi$ rescattering
effects.$^{\cite{jfd}}$  If this is the case, this effect strongly depends
on the kinematics and it is different in the off-shell nonleptonic amplitudes
entering in the calculation of the VMD diagrams. The factorization approach
provides a prediction which is free from FSI effects. In these cases we will
take these two estimates as the allowed range. On the other hand,
factorization predicts that the $\rho^0\rho^0$ intermediate state
provides most of the VMD amplitude. When both intermediate states
are taken into account in the factorization estimate, the predictions
of Eq.~(\ref{rhogam_1}) and Eq.~(\ref{rhogam_2}) roughly agree. This
will not be the case for the following mode.

\paragraph{$D^0\to\phi\gamma$: } Now, there is only one
nonleptonic intermediate state, $\phi\rho^0$. The amplitudes as
extracted from $D^0\to\phi\rho^0$ data are
\begin{eqnarray}
{\cal A}^{\rm pv}_{\rm VMD}
& = & \frac{e/f_\rho}{e/f_\phi} ~
{\cal A}_{\rm VMD}^{\rm pv}\big|_{D^0 \to \rho^0 \gamma}
 = 2.1 \times 10^{-8}~{\rm GeV}^{-1} \ \ , \nonumber\\
{\cal A}^{\rm pc}_{\rm VMD} & = & 3.5 \times 10^{-8}~{\rm GeV}^{-1}
\label{phigam_1}\ \ .
\end{eqnarray}

\noindent On the other hand, factorization predicts the much smaller
amplitudes
\begin{equation}
{\cal A}^{\rm pv}_{\rm VMD} = 0.7\times 10^{-8}~{\rm GeV}^{-1} \quad
{\rm and} \quad
{\cal A}^{\rm pc}_{\rm VMD} = 0.6\times 10^{-8}~{\rm GeV}^{-1}
\label{phigam_2}.
\end{equation}
Part of the difference between the predictions in
Eq.~(\ref{phigam_1}) and Eq.~(\ref{phigam_2}) may be due to the presence
of FSI effects in the on-shell amplitude measured and used in
Eq.~(\ref{phigam_1}) and the assumed absence of FSI in Eq.~(\ref{phigam_2}).
In Table~10 we include both predictions as the allowed range.

\paragraph{$D^0\to\omega\gamma$: }This mode is very similar to
$D^0\to\rho^0\gamma$ and we obtain from the factorization approach
of Meth.~2
\begin{equation}
{\cal A}^{\rm pv}_{\rm VMD} = 0.7\times 10^{-8}~{\rm GeV}^{-1} \qquad
{\rm and} \qquad
{\cal A}^{\rm pc}_{\rm VMD} = 0.6\times 10^{-8}~{\rm GeV}^{-1} \ \ .
\end{equation}

\paragraph{$D^+ \ra \rho^+ \gamma$:} Here, the mode $D^+ \to \rho^+ \rho^0$
should give the dominant contribution to the VMD amplitude, implying
the Meth.~2 amplitudes
\begin{equation}
{\cal A}^{\rm pc}_{\rm VMD} = 1.9\times 10^{-8}~{\rm GeV}^{-1}
\quad {\rm and} \quad
{\cal A}^{\rm pv}_{\rm VMD} = 1.6\times 10^{-8}~{\rm GeV}^{-1}\ .
\end{equation}
Incidentally, the expectation for $D^+ \ra \rho^+ \rho^0$ is
that its branching ratio should be at least 0.4\%.

\paragraph{$D^+_s \ra K^{*+} \gamma$:} Proceding analogously, we use the
factorization estimate of $D^+_s \ra K^{*+} \rho^0$ to express the VMD
amplitudes for $D^+_s \ra K^{*+} \gamma$ as
\begin{eqnarray}
{\cal A}^{\rm pv}_{\rm VMD}  &=&  \frac{e}{f_\rho}
\frac{G_F}{\sqrt{2}} \frac{a_2}{\sqrt{2}}
(m_{D_s} + m_K*) {m_\rho f_\rho \over A_1} {m_{D_s} E_\gamma} \ ,\\ \nonumber
{\cal A}^{\rm pc}_{\rm VMD} & = & \frac{m^3_{D_{s}} E_\gamma }{m_{K^*} m_\rho}
{\cal A}^{\rm pv}_{\rm VMD} \cdot \frac{2}{(m_{D_{s}}+ m_{K^*})^2}
{V\over A_1} \ \ .
\end{eqnarray}
Upon taking the form factors to be identical to those in $D^+ \ra
\bar{K}^{*0}$ in the SU(3) limit, we have
\begin{equation}
{\cal A}^{\rm pv}_{\rm VMD} = 1.0\times 10^{-8}~{\rm GeV}^{-1} \qquad {\rm and}
\qquad {\cal A}^{\rm pc}_{\rm VMD} = 0.9\times 10^{-8}~{\rm GeV}^{-1} \ \ .
\end{equation}
\phantom{xxxx}\vspace{0.15in}

Our VMD predictions for the magnitudes of the Cabibbo-suppressed
transition amplitudes are summarized in Table~10.
\phantom{xxxx}\vspace{0.1in}
\begin{center}
\begin{tabular}{l|cc}
\multicolumn{3}{c}{Table~10 {Cabibbo-suppressed VMD Amplitudes}}\\
\hline\hline
Mode & \multicolumn{2}{c}{$|{\cal A}_{\rm VMD}|~(10^{-8}~{\rm GeV}^{-1})$}
\\ \hline
& Parity-conserving & Parity-violating \\ \hline
$D^+ \to \rho^+\gamma$ & $1.6$ &  $1.9$ \\
$D_s^+ \to K^{*+}\gamma$ & $0.9$ & $1.0$ \\
$D^0 \to \rho^0\gamma$ & $ 0.2-1.0 $ &  $0.5-1.0$ \\
$D^0 \to \omega^0\gamma$ & $ 0.6 $ &  $0.7$ \\
$D^0 \to \phi^0\gamma$ & $0.6-3.5$ &  $0.9-2.1$ \\ \hline\hline
\end{tabular}
\end{center}

\begin{center}
{\bf Doubly Cabibbo-Suppressed Modes}
\end{center}

Finally, to estimate the size of the doubly Cabibbo-suppressed
modes, we consider the $D^+ \ra K^{* +} \gamma$ transition.  Upon computing
the amplitudes using the factorization expression of Eq.~(\ref{bsw_rad}),
we obtain
\begin{equation}
{\cal A}_{\rm VMD}^{\rm pv}= 4.2\times 10^{-9}~{\rm GeV}^{-1} \qquad
{\rm and} \qquad {\cal A}_{\rm VMD}^{\rm pc}= 4.4\times
10^{-9}~{\rm GeV}^{-1} \ .
\label{dcsd1}
\end{equation}
Similarly, we find for the mode $D^0 \ra K^{* 0} \gamma$
\begin{equation}
{\cal A}_{\rm VMD}^{\rm pv}= 1.75 \times 10^{-9}~{\rm GeV}^{-1} \qquad
{\rm and} \qquad {\cal A}_{\rm VMD}^{\rm pc}= 1.83 \times
10^{-9}~{\rm GeV}^{-1} \ .
\label{dcsd2}
\end{equation}

\section{\bf Summary and Conclusions}
The existing database for direct evidence of radiative $D$ decays is
meagre, as has been shown in Table~1. However, interesting levels of
experimental sensitivity are currently being attained and we can
anticipate the detection of radiative signals in the not-too-distant
future. Our motivation in undertaking the study reported here has
been to stimulate such experimental efforts.

As shown here, weak radiative decays of
charmed mesons are not dominated by the short distance penguin
diagrams of Figs.~1(a),1(c), but rather by long distance processes
involving nonperturbative strong interaction dynamics.  From the standpoint
of probing the inner workings of the Standard Model, one might have naively
hoped to use radiative charm decays in order to observe the short distance
process $c\to u\gamma$.  This would be in analogy with radiative $B$-meson
decay, where the amplitude is dominated by the penguin transition
$b\to s\gamma$ and receives a large enhancement from QCD radiative
corrections. By contrast, the corresponding $c\to u\gamma$ charm transition
is minuscule at lowest order and would require an unexpectedly large QCD
enhancement to become detectable.  To our knowledge, the calculation
performed here of the QCD radiative correction to $c\to u\gamma$ is the
first explicit and detailed analysis of this system given in the
literature. In addition, we were able to employ $U$-spin arguments to
clarify the role played by neutral, flavor-changing operators such as
$\bar{u}c\bar{q}q$ which contribute to the expanded operator basis in the
$RG$ analysis.  Our conclusion that the $c\to u\gamma$ QCD radiative
corrections are substantially larger than for $b\to s\gamma$ is due in
part to the large operator mixing at the lower renormalization scale
associated with the $c$-quark and in part to the disparate sizes of the
Wilson coefficients at the matching scale of the contributing operators.
Nevertheless, the radiatively-corrected $c\to u\gamma$
penguin transition remains extremely small.  The main sources of
suppression are the small quark masses and also the CKM factors
$|V_{cb}^{*}V_{ub}|^2$ occurring in the numerator of the
$c\to u\gamma$ branching fraction in Eq.~(\ref{qcdchbr}).
In the Wolfenstein parameterization, with $\sin\theta_c=\lambda$,
this CKM dependence amounts to a $\lambda^{10}$ suppression in decay rate.

On the other hand, we have shown in Sections~3 and~4 that long
distance contributions are several orders of magnitude larger.
A very rough estimate of the typical branching ratio to be expected is
\beq
B_{D\to M\gamma}\sim  \alpha_{\rm em} B_{D\to M}^{\rm non-lept.}  \ ,
\label{esti}
\eeq
where $B_{D\to M}^{\rm non-lept.}$ is the branching ratio for the
nonleptonic $D$ decay to some final state $M$. Thus, typical branching
ratios of order $B_{D\to M}^{\rm non-lept.}\sim (0.001-0.05)$ would
induce radiative branching ratios in the range $B_{D\to M\gamma}\sim
(7\times 10^{-6}\to 4\times 10^{-4})$.  In Sections~3 and~4 we
have performed a more detailed analysis by modeling the nonperturbative
dynamics.

Inspecting the long distance contributions to the set of exclusive processes
$D^0\to\rho\gamma$, $D^0\to\omega\gamma$, $D_{s}\to K^{*+}\gamma$ and
$D^+\to\rho^+\gamma$ for which $c\to u\gamma$ is the underlying
transition, we see that the VMD and pole amplitudes carry a single factor
$\lambda$, therefore representing an
enhancement of $\lambda^4$ over the penguin amplitude.
As can be seen in Table~11 the expected branching fractions for
these modes are in the $10^{-6}-10^{-4}$ range, whereas we estimate
$B_{c\to u\gamma}\sim 10^{-12}$.
As a consequence, $c\to u\gamma$ is not a good process to test the
validity of the Standard Model. That is, a hypothetical contribution
from new physics would have to be extremely large in order to overcome
the long distance physics.

The situation is very different in radiative $B$ decays. The short
distance transition $b\to s\gamma$ has the same CKM structure as the
corresponding long distance contributions. For instance, the mode
$B\to K^*\gamma$ might conceivably have long distance contamination
of the order of $20\%$ in the
rate $^{\cite{gp}}$. Although this is small compared to the
charm case,
 it would be desirable to reduce the uncertainty in the calculation
of these
effects in order to subtract them from the measured signal.
Moreover, long
distance effects could also be affecting the inclusive $b\to s\gamma$
branching ratio, therefore limiting the precision with which the
Standard Model can be tested in these decays.

The various amplitudes are summarized in Table~11 and are given there
in units of $10^{-8}~{\rm GeV}^{-1}$.  In principle, the most
conservative attitude is to take all relative signs as unknown, which
would render the calculation of branching ratios highly uncertain.
Fortunately, with the aid of the quark model we can reduce this ambiguity.
The relative sign of Pole-II to Pole-I contributions is affected by
(i) a minus sign difference in the pole denominators, (ii) an extra
minus sign in type-II amplitudes due to the vector meson propagator, and
(iii) minus sign differences in the $VP\gamma$ couplings between the
$c$-quark and light-quark EM sectors.  In a quark description of a
$q_1 {\bar q}_2$ meson, this latter sign is inferred by studying
\begin{equation}
h_{VP\gamma}= e \left(\frac{q_1}{m_1} + \frac{q_2}{m_2} \right) \ \ .
\label{mu_qm}
\end{equation}
Although this line of reasoning narrows down the range of
predictions significantly, experimental data will be needed
to obtain information regarding the relative phase between the
pole and VMD contributions.  In this regard, it will be helpful
to note that, at least in our approach, the parity-violating
amplitudes arise solely from the VMD process.
\phantom{xxxx}\vspace{0.1in}
\begin{center}
\begin{tabular}{l||ccc|c|c}
\multicolumn{6}{c}{Table~11 {Amplitude and Branching Fraction
Predictions}}\\
 \hline\hline
Mode & \multicolumn{3}{c}{${\cal A}^{\rm pc}$} &
${\cal A}^{\rm pv}$ & $B_{D\to M\gamma}$ ($10^{-5}$) \\ \hline
                                 & P-I & P-II & VMD & VMD &  \\ \hline
$D^{+}_{s} \rho^+\gamma$ & $8.2$ & $-1.9$ & $\pm 3.2$ & $\pm 2.8$ & $6 - 38$\\
$D^0 {\bar K}^{*0}\gamma$ & $5.6$ & $-5.9$ & $\pm 3.8$ & $\pm (5.1-6.8)$ &
$7 - 12$ \\
$D^{+}_{s} b_1^+\gamma$ & $7.2$ & --- & --- & --- & $\sim 6.3$\\
$D^{+}_{s} a_1^+\gamma$ & $1.2$ & --- & --- & --- & $\sim 0.2$\\
$D^{+}_{s} a_2^+\gamma$ & $2.1$ & --- & --- & --- & $\sim 0.01$\\
$D^+ \rho^+\gamma$ & $1.3$ & $-0.4$ & $\pm 1.6$ & $\pm 1.9$ & $2 - 6$ \\
$D^{+} b_1^+\gamma$ & $1.2$ & --- & --- & --- & $\sim 3.5$\\
$D^{+} a_1^+\gamma$ & $0.5$ & --- & --- & --- & $\sim 0.04$\\
$D^{+} a_2^+\gamma$ & $3.4$ & --- & --- & --- & $\sim 0.03$\\
$D_s^+ K^{*+}\gamma$ & $2.8$ & $-0.5$ & $\pm 0.9$ & $\pm 1.0$ & $0.8 - 3$ \\
$D^{+}_{s} K_2^{*+}\gamma$ & $6.0$ & --- & --- & --- & $\sim 0.2$\\
$D^0 \rho^0\gamma$ & $0.5$ & $-0.5$ & $\pm (0.2 - 1.0)$ & $\pm (0.6 - 1.0)$ &
$0.1-0.5$\\
$D^0 \omega^0\gamma$ & $0.6$ & $-0.7$ & $\pm 0.6$ & $\pm 0.7$ & $\simeq 0.2$ \\
$D^0 \phi^0\gamma$ & $0.7$ & $-1.6$ & $\pm ( 0.6 - 3.5)$ & $\pm (0.9 - 2.1)$ &
$0.1 - 3.4$ \\
$D^{+} K^{*+}\gamma$ & $0.4$ & $-0.1$ & $\pm 0.4$ & $\pm 0.4$ & $0.1 - 0.3$\\
$D^{0} K^{*0}\gamma$ & $0.2$ & $-0.3$ & $\pm 0.2$ & $\pm 0.2$ & $\simeq 0.01$\\
\hline\hline
\end{tabular}
\end{center}
\phantom{xxxx}\vspace{0.15in}

Finally, let us comment on the inclusive photon spectrum.
In the $B$ system, the quark transition $b\to s\gamma$
provides a useful framework for predicting properties of the hadronic
inclusivedecay $B\to X_s \gamma$.  Thus, one estimates the $B\to X_s \gamma$
decay rate by computing the $b\to s\gamma$ decay rate and normalizing
relative to the semileptonic decays to eliminate undue dependence on the mass
$m_b$.  Likewise, one predicts the photon energy spectrum in $B\to X_s
\gamma$ decay by referring to the underlying two-body $b\to s\gamma$
decay.$^{\cite{fnl1},\cite{fnl2}}$  If quarks were
free, there would be a monochromatic photon spike at $E_\gamma =
(m_b^2 - m_s^2 )/2m_b \simeq m_b/2$.
In reality, the photon spectrum becomes broadened via hadronization
of the
$s$-quark jet.  The individual strange mesons ($K^*(892)$,
$K_1(1270)$, {\it
etc}) which populate the inclusive final state $X_s$ originate
predominantly
from the $s$-quark jet hadronization.  These explanations of $B\to
X_s \gamma$
inclusive decay are in accordance with the spectator model, and so
isospin symmetry should manifest itself event-by-event.  For example,
the rates for isospin-related modes such as $B^0 \to K^{*0}\gamma$
and
$B^- \to K^{*-}\gamma$ should be equal.  A deviation from this
pattern would constitute evidence for either non-spectator or
new-physics
contributions.  In a heavy-quark effective theory description,
such non-spectator effects would occur at subleading level.

The theoretical description of charm inclusive decay could hardly be
more different.  Now, there is no emergent light-quark jet which
hadronizes to form the set of final states.  Instead, the `black box'
of long-range effects such as pole-amplitudes, VMD-amplitudes,
{\it etc} dominates the physics.  Thus, to determine the photon
energy
spectrum in $D\to X_u \gamma$, one would sum over the most important
of the exclusive radiative modes.  Presumably this would yield a
reasonable description at least over the part of phase space where
the photon energy is largest.  It would be prudent to be on the
lookout
for the unexpected.  For example, exclusive modes in light meson
radiative
decay are known to exhibit rather large isospin-violating effects, as
in
\beq
{\Gamma_{K^{*0}\to K^0\gamma} \over \Gamma_{K^{*+}\to K^+\gamma}}
= 2.27 \pm 0.30 \quad {\rm and} \quad
{\Gamma_{\rho^{0}\to \pi^0\gamma} \over \Gamma_{\rho^{+}\to
\pi^+\gamma}}
= 1.76 \pm 0.49 \ .
\label{c1}
\eeq
If this effect were to be maintained mode by mode in the exclusive $D$
decays, it would lead to interesting levels of isospin violation in
the inclusive decay.  Of our results, Table~11 indicates that the
likeliest possibility for isospin violation would appear to be in
$D^0\to \rho^0\gamma /D^+\to \rho^+\gamma$.

Charm radiative decays give us the opportunity to study various
aspects of long distance dynamics. We have seen that the theoretical
predictions of the branching fractions are, in some cases, rather
uncertain due to model dependence.  Experimental information will
therefore be needed to complete the theoretical picture of these decays.
It is an interesting irony that the understanding gained from
future observation of different $D$ radiative decays can
then be used to predict more confidently the size of such effects in
$B$ decays.

The research described in this paper was supported in part by
the National Science Foundation and the Department of Energy.
We gratefully acknowledge useful conversations with T.E. Browder,
A. Datta, T. Rizzo, X. Tata and S. Willenbrock.
We also wish to thank K.J. Moriarity for his assistance with
numerical calculations.

\appendix{\bf Appendix: VMD in Light-Meson Radiative Decay}

The original application of VMD for analyzing hadronic radiative decays
occurred in the light meson sector.$^{\cite{gmsw}}$  In order to test the
VMD method using an up-to-date database, we too shall consider (briefly)
light meson radiative decays in this Appendix.  As we shall see from our
study of two particularly clean examples, the situation is encouraging
but not uniformly so.  First, we shall revisit the original arena for
testing VMD, the $\rho$ and $\omega$ decays into pion-photon final
states.  Then we shall analyze decays of a higher mass state, the
tensor meson $A_2^+ (1320)$.  We stress that in each of these cases
the transition is purely electromagnetic, unlike the more complicated
electroweak decays treated in the main body of the paper.  Therefore,
the `pole' amplitudes do not occur here since there is no weak mixing,
so one obtains a clean look at the VMD contribution.

\begin{center}
{\bf Radiative Decays of the Vector Mesons $\rho$ and $\omega$}
\end{center}

There are three electromagnetic $P$-wave decays in the $\rho$ -- $\omega$
system,
\beq
\omega \to \pi^0\gamma~, \qquad \rho^+ \to \pi^+ \gamma \quad {\rm and}
\quad \rho^0 \to \pi^0 \gamma \ .
\label{a10}
\eeq
In the VMD approach, these are described in terms of two
electromagnetic mixing amplitudes, $\omega$--$\gamma$ and
$\rho$--$\gamma$, and one strong interaction vertex,
$g_{\omega\rho\pi}$.

Due to the off-shell nature of the VMD amplitudes, different momentum
regions occur in the $\omega\rho\pi$ vertex for the transitions of
Eq.~(\ref{a10}).  In $\omega \to
\pi\gamma$, the intermediate $\rho$ propagates at $q^2 = 0$
whereas for $\rho \to \pi\gamma$ it is the intermediate
$\omega$ which propagates at $q^2 = 0$.  Part of the VMD folklore
built up over the years is that extrapolation of the light vector
meson squared-momenta from the meson mass-shell to the photon mass-shell
does not strongly affect the decay amplitude.  The ratio of $\rho$
and $\omega$ decay widths can be used to test this as follows.  Recall that
for the VMD description of $1^- \to 0^- \gamma$ transitions, the strong vertex
$g_{\omega\rho\pi}$ is related to the decay width $\Gamma$ via
\beq
g_{\omega\rho\pi} = \left[ {f\over e}^2 \cdot
{12\pi\Gamma\over |{\bf q}|^3} \right]^{1/2} \ \ ,
\label{a11}
\eeq
where $f \to f_\rho$ for $\omega$ decay and $f \to f_\omega$ for
$\rho$ decay.  Noting that the decay momenta in $\omega \to
\pi\gamma$ and $\rho \to \pi\gamma$ are almost equal, one has
\beq
\bigg|{f_\omega \over f_\rho}\bigg| = \sqrt{ \Gamma_{\omega\to\pi^0\gamma}
\over \Gamma_{\rho^+\to\pi^+\gamma}} = 3.24 \pm 0.19 \ ,
\label{a12}
\eeq
provided the {\it same} strong vertex is used in each decay.  In the
above, we have used the charged-$\rho$ decay width in view of its superior
accuracy.  The value appearing in Eq.~(\ref{a12}) is seen to be in
accord with that inferred from vector meson decay into lepton pairs
({\it cf} Table~1 of Ref.~\cite{gp}),
\beq
\bigg|{f_\omega \over f_\rho}\bigg| = 3.39 \pm 0.10 \ \ .
\label{a13}
\eeq

Alternatively, one can use each of these radiative decays to
extract determinations of $g_{\omega\rho\pi}$ as in Eq.~(\ref{a11}),
and one finds
\beq
g_{\omega\rho\pi}= \left\{ \begin{array}{ll}
(11.73 \pm 0.35)~{\rm GeV}^{-1} & (\omega \to \pi^0\gamma)\\
(12.40 \pm 0.64)~{\rm GeV}^{-1} & (\rho^+ \to \pi^+ \gamma) \\
(16.40 \pm 2.1)~{\rm GeV}^{-1} & (\rho^0 \to \pi^0 \gamma) \ \ .
\end{array}\right.
\label{a14}
\eeq
The $\omega \to \pi^0\gamma$ and $\rho^+ \to \pi^+ \gamma$
determinations are seen to be consistent within experimental
error.  This is significant because these decays involve
different momentum extrapolations as discussed above.  The
larger coupling obtained from $\rho^0 \to \pi^0 \gamma$ decay
has substantially larger errors.  We now turn to a different
transition in which, if one accepts the data at face value,
a non-negligible momentum dependence is present.

\begin{center}
{\bf Decays of the Tensor Meson $A_2^+ (1320)$}
\end{center}
The meson $A_2^+ (1320)$ has been observed to decay into both
the $\pi \rho$ and $\pi \gamma$ modes, with branching ratios
\beq
{\rm B}_{A_2 \to \pi\rho} = 0.701 \pm 0.027 \qquad {\rm and} \qquad
{\rm B}_{A_2 \to \pi\gamma} = (2.8 \pm 0.6)\cdot 10^{-3} \ \ .
\label{a1}
\eeq
These data turn out to provide a particularly clean test of the
VMD method in two respects.  First, there is just a single partial
wave in the final state.  As a consequence, the decay rates alone
can be used to test VMD without any need for polarization information
of the final state particles.  The occurrence of a single orbital
angular momentum in the final state follows from conservation of
parity and of angular momentum. Thus we have
\beqa
{\cal P}:\qquad + &=& (-)^2 (-)^L \ \Longrightarrow \ L = 0,2,4,\dots \\
{}~{\cal J}:\qquad  |{\bf 2}| &=& |{\bf 1} + {\bf L}| \ \ ~
\ \Longrightarrow \ L = 1,2,3
\eeqa
which implies that $L=2$.  In addition, of the three light vector
mesons $\rho, \omega, \phi$, only the $\rho$ can appear together
with a pion in a final state of $A_2$ decay.  The reason is that
the decay $A_2 \to \pi ~V$ ($V$ is a vector meson) proceeds through
the strong interactions and conservation of $G$-parity forbids the
$\pi\omega$ and $\pi\phi$ modes.  Thus, the rho is the only light
vector meson involved in the VMD determination and interference with
$\omega$ or $\phi$ mediated processes is absent.

The amplitude for the transition $A_2^+ (p)\to \pi(q)^+ \rho^0(k)$
can be written as
\beq
{\cal A}_{\pi\rho} = {g_{\pi\rho}\over m_A^2}\epsilon^{\mu\nu\alpha\beta}
p_\mu \epsilon^\dagger_\nu (k) q_\alpha h_{\beta\sigma}(p) q^\sigma \ ,
\label{a3}
\eeq
where $g_{\pi\rho}$ is a dimensionless quantity and
$h_{\beta\sigma}(p)$ is the spin-two polarization tensor of the $A_2$.
{}From the decay rate relation,
\beq
\Gamma_{A_2^+ \to \pi^+\rho^0} = {g^2_{\pi\rho}\over 40\pi}
{{\bf q}_{\pi\rho}^5 \over m_A^4} \ \ ,
\label{a4}
\eeq
one determines a magnitude for the coupling $g_{\pi\rho}$.  This
can be used, in turn, to predict the radiative coupling
$g_{\pi\gamma}$ via the VMD formula
\beq
g^{\rm VMD}_{\pi\gamma} = {e\over f_\rho} g_{\pi\rho} \ \ ,
\label{a5}
\eeq
and we find
\beq
g^{\rm VMD}_{\pi\gamma} = 1.99 \pm 0.06 \ \ .
\label{a6}
\eeq

Alternatively, it is possible to determine the pion-photon coupling
{\it directly}.  Analogous to Eq.~(\ref{a3}), we can write down
a gauge-invariant photon-emission transition,
\beq
{\cal A}_{\pi\gamma} = {g_{\pi\gamma}\over
m_A^2}\epsilon^{\mu\nu\alpha\beta}p_\mu \epsilon^\dagger_\nu (k)
q_\alpha h_{\beta\sigma}q^\sigma \ .
\label{a7}
\eeq
Fixing the coupling $g_{\pi\gamma}$ in terms of the decay rate
\beq
\Gamma_{A_2^+ \to \pi^+\gamma} = {g^2_{\pi\gamma}\over 40\pi}
{{\bf q}_{\pi\gamma}^5 \over m_A^4} \ \ ,
\label{a8}
\eeq
yields the value
\beq
g^{\rm expt}_{\pi\gamma} = 0.98 \pm 0.11 \ \ .
\label{a9}
\eeq

Thus, one obtains a factor-of-2 discrepancy between the
empirical amplitude and the VMD prediction, with the VMD value
being the larger.  Several possible explanations for the lack
of agreement come to mind.  Although the radiative branching ratio
given in Eq.~(\ref{a1}) has reasonably small error bars, the
signal is based on only one experiment.  Alternatively, there may be
unexpectedly large momentum dependence in the $A_2\pi\rho$ vertex.
Thus, as one proceeds from the rho mass-shell ($k^2 = m_\rho^2$) to
the photon mass-shell ($k^2 = 0$), a 'softening' might occur in
the VMD estimate.  However, to our knowledge there is no previous
evidence for such momentum dependence for the $\rho$ extrapolation.

\end{document}